\documentclass[journal]{new-aiaa}
\usepackage{new-mysymb}
\usepackage{longtable}
\usepackage{multirow}
\usepackage{pbox}
\usepackage[hyphens]{url}
\usepackage{xcolor}
\usepackage{soul}
\usepackage{nomencl}
\usepackage{booktabs}
\usepackage{etoolbox}
\usepackage{adjustbox}
\usepackage[ruled,vlined,linesnumbered]{algorithm2e}

\DeclareMathOperator*{\argmin}{arg\,min}

\newcommand\GP{\mathcal{GP}}

\newcommand\bE{\mathbb{E}}
\newcommand\bR{\mathbb{R}}

\newcommand\cD{\mathcal{D}}

\newcommand\cG{\mathcal{G}}

\newcommand\cL{\mathcal{L}}
\newcommand\cN{\mathcal{N}}
\newcommand\cR{\mathcal{R}}

\soulregister\cite7
\soulregister\ref7

\newcommand\bX{\mathbb{X}}

\newcommand\vtvf{\boldsymbol{\varphi}}

\usepackage[utf8]{inputenc}
\usepackage{makecell}
\usepackage{float}
\usepackage{array}
\usepackage{graphicx}
\usepackage[version=4]{mhchem}
\usepackage{longtable,tabularx}
\usepackage{amsmath}

\title{Modal Analysis of Spatiotemporal Data \\ via Multivariate Gaussian Process Regression}

\author{Jiwoo Song\footnote{Graduate Student Research Assistant, Student Member AIAA, jzs6565@psu.edu.} and Daning Huang\footnote{Assistant Professor, Member AIAA, daning@psu.edu.}}
\affil{Department of Aerospace Engineering, The Pennsylvania State University, University Park, PA, 16802}

\begin{document}

\maketitle

\begin{abstract}
Modal analysis has become an essential tool to understand the coherent structure of complex flows.
The classical modal analysis methods, such as dynamic mode decomposition (DMD) and spectral proper orthogonal decomposition (SPOD), rely on a sufficient amount of data that is regularly sampled in time.
However, often one needs to deal with sparse temporally irregular data, e.g., due to experimental measurements and simulation algorithm.
To overcome the limitations of data scarcity and irregular sampling, we propose a novel modal analysis technique using multi-variate Gaussian process regression (MVGPR).
We first establish the connection between MVGPR and the existing modal analysis techniques, DMD and SPOD, from a linear system identification perspective.
Next, leveraging this connection, we develop a MVGPR-based modal analysis technique that addresses the aforementioned limitations. The capability of MVGPR is endowed by its judiciously designed kernel structure for correlation function, that is derived from the assumed linear dynamics.
Subsequently, the proposed MVGPR method is benchmarked against DMD and SPOD on a range of examples, from academic and synthesized data to unsteady airfoil aerodynamics.  The results demonstrate MVGPR as a promising alternative to classical modal analysis methods, especially in the scenario of scarce and temporally irregular data.
\end{abstract}

\section*{Nomenclature}

{\renewcommand\arraystretch{1.0}
\noindent\begin{longtable*}{@{}l @{\quad=\quad} l@{}}
$\vC$ & Covariance \\ 
$\cD$ & Dataset \\
$d(\bullet,\bullet)$ & Grassmannian distance \\
$\bE$ & Expectation operator \\
$\mathcal{GP}$ & Gaussian process \\
$g, \mathbf{g}$ & Observable \\
$\mathcal{K}, \mathbb{K}$ & Koopman operator \\
$k(t, t'), \mathbf{K}(t, t')$ & Kernel function or correlation function of a Gaussian process \\
$\vK_{\text{LMC}}(\tau, \vg_0', \vg_0'')$ & LMC kernel of MVGPR \\
$\cN$ & Gaussian distribution \\
$\mathbf{S}(\omega)$ & Power spectral density function \\
$\vU$, $\mathbf{U}_r$ & Left singular vector matrix \\
$\vV$, $\mathbf{V}_r$ & Right singular vector matrix \\
$\lambda, \mathbf{\Lambda}$ & Eigenvalues \\
$\xi$ & Spatial coordinate of flow field \\
$\vtS$ & Singular value matrix \\
$\sigma$ & Standard deviation, Singular value \\
$\vtvf$ & Eigenfunction of SPOD\\
$\vtf$ & Modal coordinates\\
$\vtY$ & Left eigenvector matrix\\
$\omega$ & Angular frequency \\

\end{longtable*}}

\section{Introduction}

Spatiotemporal data is universally seen in aerospace and mechanical engineering applications, especially in fluid and structural dynamics.
One fundamental step to understand the underlying physical phenomena is to extract the dominant modes and the associated characteristic frequencies from the spatiotemporal data.
For example, proper orthogonal decomposition (POD) \cite{berkooz1993proper} and spectral POD (SPOD) \cite{schmidt2020guide, towne2018spectral} methods have been used to extract out coherent flow structures from turbulent flow; dynamic mode decomposition (DMD) \cite{schmid2011applications,kutz2016dynamic} has been employed to produce reduced-order dynamical systems representing the evolution of complex flow fields.  
Furthermore, due to its connection to the more general Koopman mode decomposition (KMD) techniques \cite{brunton2021modern}, the DMD-type techniques have enjoyed some prevalence in data-driven modal analysis of spatiotemporal data.
The DMD has been further extended with the resolvent analysis techniques \cite{taira2017modal, taira2020modal} to identify potential forcing and response modes in the flow field that may reveal potential flow mechanisms hidden in the spatiotemporal data.
Note that, for DMD, when multiple trajectories need to be considered for either noise filtering or capturing multiple modes, one can combine them using the so-called ensemble DMD \cite{tu2013dynamic}. In this paper our primary focus is on the multi-trajectory case, and hence DMD is considered in an ensemble setting, unless otherwise specified.

The DMD and SPOD methods have been widely studied in literature, and their connections and distinctions have been clarified  (see, e.g., \cite{towne2018spectral}). POD method decomposes the stochastic ensemble dataset into deterministic modes, effectively capturing predominant energy or variance within the data. This optimality of the method stems from the spatial orthogonality of the modes, leading to its alternative nomenclature, space-only POD. In spectral POD, the orthogonality is extended to the spatiotemporal domain, and the orthogonal modes within the same frequency spectrum are obtained. This feature of SPOD makes it optimal at representing spatiotemporal coherence of the data \cite{schmidt2020guide}.  In this perspective, for stationary flows resulting from stochastic processes, DMD is nearly equivalent to SPOD as it extracts the modes in which the latent processes evolve in harmonics.  Alternatively, one can view SPOD as a combination of DMD and POD, incorporating the temporal correlation and statistical optimality, respectively.  This perception of considering flow field as a stochastic process further extends its applicability to the more complex problem. In Ref. \cite{arbabi2017study}, the analysis of post-transient flow is performed using KMD.  In their work, they observed the connection between the power spectral density of a stochastic process -- inverse Fourier transform of covariance of the process -- and the concept of Koopman spectral density.  Utilizing this connection, they identified the Koopman continuous and discrete spectra using spectral estimation techniques.

Although DMD-type algorithms appeal to Koopman spectral density of the system, they are known to be sensitive to sensor noise.  To overcome this issue, other variants of DMD such as total-least-squares DMD \cite{hemati2017biasing}, residual DMD \cite{colbrook2021rigorous}, and sparsity-promoting DMD \cite{pan2021sparsity} were developed and shown to be superior to the regular DMD method in robustness to the noise and spectral pollution.  However, it is still a deterministic model and cannot consider the statistics of the data, which motivated the development of probabilistic models \cite{takeishi2017bayesian, kawashima2023gaussian}. Bayesian DMD, one of the variants of DMD method, formulates the DMD algorithm in Bayesian framework \cite{takeishi2017bayesian}. Therefore, the model has a capability of providing a treatment on system uncertainty along with the accurate estimation of eigenvalues that is comparable to total-least-squares DMD. As an extension of Bayesian DMD algorithm, there was an explicit attempt to combine Gaussian process and KMD, which they called Gaussian Process Koopman Mode Decomposition (GPKMD) \cite{kawashima2023gaussian}. The GPKMD model is a multi-variate probabilistic generative model of KMD. As the name implies, it induces the correlation between the state variables via Koopman modes.  While the model makes use of kernel functions to construct the covariance matrix of data, they did not assume any particular form of dynamics in the system, which is potentially inappropriate for modal analysis.

In spite of their prevalent usage, current modal analysis techniques encounter some challenges. Some of them include (1) Spectral pollution caused by discretization of the dynamics as well as the measurement noise \cite{colbrook2021rigorous}, (2) Inconvenience to utilize temporally non-uniformly distributed data, due to the usual underlying assumption of uniform time step size in conventional DMD methods \cite{colbrook2021rigorous}. This work primarily focuses on addressing the latter challenge. To provide context to the issue, acquiring data with uniformly distributed time steps throughout an entire experimental period can be difficult. Recent research, such as the study on identifying SPOD modes using DMD for nonsequential pairwise data \cite{cardinale2022spectral}, addresses the temporal irregularity. Another scenario is a flow simulation with adaptive time stepping, where the solutions are saved with non-uniform time steps \cite{henry2023pelec}.
To tackle this challenge, we first note a simple fact from linear system theory \cite{edition2002probability}. For a single-input single-output (SISO) system, if the input is Gaussian white noise, the output is a zero-mean Gaussian process (GP) with its correlation function determined by the system dynamics. Similarly, if the samples at initial time follow Gaussian distribution, the corresponding initial condition response is also a GP. Such relation extends to higher-dimensional dynamical systems \cite{sarkka2013spatiotemporal}. In other words, the identification of a linear system, of which DMD is a special case, and the estimation of a GP correlation function are equivalent.

In this work, we extend multi-variate Gaussian process regression (MVGPR) to develop a novel modal analysis approach for spatiotemporal data, while establishing the connection between MVGPR and the DMD and SPOD methods.  Specifically, MVGPR is shown to produce a low-order linear dynamical system for a spatiotemporal dataset. Unlike the GPKMD model, when the correlation function of MVGPR is designed with a particular structure, the model will only admit a sparse spectrum, and thus has the potential to alleviate spectral pollution.  Assuming stationarity of the data in time, the learning of correlation function only depends on the differences between time steps and the measurements at that time instances, without requiring the time step size to be uniform, which will be potentially more advantageous over DMD and SPOD methods. Finally, to make the MVGPR approach more general, we establish the connection between MVGPR and Koopman operator theory, and show that the MVGPR models can also learn from nonlinear dynamical data.

Building on the general-purpose MVGPR framework developed previously by the authors \cite{Sadagopan2021a,Sadagopan2022, song2023modal}, this study aims to demonstrate the feasibility of GPR-based modal analysis of spatiotemporal data, as an alternative to DMD and SPOD methods.  Specifically, the goals of this paper are to
\begin{enumerate}
    \item Present the theoretical formulation of MVGPR for modal analysis of spatiotemporal data.
    \item Reveal the connections among MVGPR, linear system identification, Koopman analysis via DMD, and SPOD.
    \item Benchmark the MVGPR against DMD and SPOD on multiple problems to highlight its robustness in the presence of temporally irregular data.
\end{enumerate}

The remainder of the paper is organized as follows. In Sec. \ref{sec2}, the preliminaries for the Koopman operator, (Extended) DMD, and MVGPR algorithms are presented. In Sec. \ref{sec3}, the identification method of stationary flow system using MVGPR will be discussed. Also, we will demonstrate the connection between DMD, SPOD, and MVGPR methods. In Sec. \ref{sec4}, the results of the different modal analysis methods are shown for both full measurement and temporally irregular measurement data. In Sec. \ref{sec5}, we summarize the conclusions of the paper with potential future works. 

\section{Preliminaries}\label{sec2}
As a basis for the development of new methodology, this section presents a brief summary of the Koopman analysis framework, the standard (Extended) DMD technique, and the Multivariate Gaussian Process Regression (MVGPR).

\subsection{Data-driven Koopman Mode Decomposition}

\subsubsection{Koopman Mode Analysis}
In this study, we are interested in autonomous nonlinear dynamics whose state is $\vx \in \bX \subseteq \bR^n$
\begin{equation}
    \dot\vx = \vf(\vx),\quad \vx(0)=\vx_0
\end{equation}

where dynamics, $\vf: \bX\to\bR^n$ takes the state of the system and maps to the derivative of the state.  From an operator theoretic point of view \cite{brunton2021modern,mezic2013analysis}, there exists a mapping of the finite dimensional states $\vx$ to a functional $g: \bX \to \cG(\bX)$, called observable in an infinite dimensional space $\cG(\bX)$ equipped with inner product defined in Eq. \eqref{eqn:inprod}.
\begin{equation}\label{eqn:inprod}
    \langle u, v\rangle = \int_{\bX} v^*(\vx) u(\vx) d\vx
\end{equation}
where $\Box^*$ indicates complex-conjugate. In this space, the functional evolves linearly in time with the (continuous-time) Koopman operator, $\mathcal{K}$.
\begin{equation}
    \dot g = \mathcal{K} g
\end{equation}
For nonlinear dynamics in the attraction basin of equilibrium points and periodic orbits, e.g., limit cycles, one can identify a finite set of pairs of eigenvalues $\lambda_i=\zeta_i+i\omega_i$ and eigenfunctions $\phi_i(\vx)$, such that
\begin{equation}
    \mathcal{K}\phi_i = \lambda_i\phi_i
\end{equation}
The set of eigenpairs form an invariant space, and if an observable $g_j$ can be decomposed in terms of the eigenfunctions,
\begin{equation}
    g_j(\vx) = \sum_{i=1}^N \phi_i(\vx) \langle \phi_i(\vx), g_j(\vx) \rangle
\end{equation}
then the dynamical response of the observable can be represented by a finite-dimensional series
\begin{equation}
    g_j(t) = \sum_{i=1}^N \exp(\lambda_i t)\phi_i(\vx_0) \langle \phi_i(\vx_0), g_j(\vx_0) \rangle \equiv \sum_{i=1}^N \exp(\lambda_i t)\Gamma_{ij}
\end{equation}
Suppose there are multiple observables, $\vg=[g_1,g_2,\cdots,g_M]^T$,
\begin{equation}\label{eqn_obs_dyn}
    \vg(t) = \sum_{i=1}^N \exp(\lambda_i t) \vtg_i \equiv \vtG\exp(\vtL t)
\end{equation}
where $\vtg_i=[\Gamma_{i1},\Gamma_{i1},\cdots,\Gamma_{iM}]^T$ represents a dynamic mode in $\vg$ that evolves with eigenvalue $\lambda_i$, i.e., with frequency $\omega_i$ and damping $\zeta_i$.

\subsubsection{Modal Analysis by (Extended) DMD}
Dynamic Mode Decomposition (DMD) is a standard method for extracting dynamically relevant modes, or arguably the Koopman modes, from a dataset of trajectories.  Consider a set of data snapshot pairs, $\cD=\{(t_i,\vx_i, \vy_i)\}_{i=1}^m$ sampled uniformly with time step size $\Delta t$, one may form data matrices $\mathbf{X}, \mathbf{Y} \in \mathbb{R}^{n \times m}$ with observables $\vg$,
\begin{equation}\label{data_matrix}
    \mathbf{X} = \begin{bmatrix}
    \vg(\mathbf{x}_1) & \vg(\mathbf{x}_2) & \cdots & \vg(\mathbf{x}_m)
    \end{bmatrix}, \quad \mathbf{Y} = \begin{bmatrix}
    \vg(\mathbf{y}_1) & \vg(\mathbf{y}_2) & \cdots & \vg(\mathbf{y}_m)
    \end{bmatrix}
\end{equation}
The DMD effectively produces a finite-dimensional discrete-time approximation $\vA$ of the Koopman operator, such that $\vY=\vA\vX$; standard least squares gives $\vA=\vY\vX^+$, where $\Box^+$ is pseudo-inverse.  For practical calculations where the dimension of $\vg$ is high, it is typical to approximate the pseudo-inverse by a truncated SVD,
\begin{equation}
    \mathbf{X} = \mathbf{U}\mathbf{\Sigma} \mathbf{V}^H \approx \mathbf{U}_r \mathbf{\Sigma}_r \mathbf{V}_r^H \Rightarrow \vX^+=\vV_r\vtS_r^{-1}\vU_r^H
\end{equation}
where $\Box^H$ is Hermitian tranpose and $r$ is a reduced dimension.  Next, via the eigendecomposition of a projected $\vA$,
\begin{equation}
    \bar{\vA} \equiv \vU^H_r\vY\vV_r\vtS_r^{-1} = \vW\bar{\vtL}\vS^H
\end{equation}
one can recover the eigenpairs of $\vA$, i.e., the approximate Koopman modes and Koopman eigenvalues,
\begin{equation}
    \vtG = \vY \vV_r \vtS_r^{-1} \vW,\quad \vtL = \log(\bar{\vtL})/\Delta t
\end{equation}

As mentioned earlier, the conventional DMD approach involves a potential issue. A typical technique is to employ an ensemble technique to average out DMD results over many noisy trajectories, yet such approach relies on the availability of a large high quality dataset that is often unavailable \cite{duke2012error}.  Another issue is the missing data, such that the data samples are not uniformly distributed over time.  This issue might completely overthrow the DMD formulation.  One remedy is to employ an optimization-based formulation to directly learn the continuous-time operator, the computational complexity of which, however, is high \cite{askham2018variable}.

\subsection{Multivariate Gaussian Process Regression}

In the context of dynamical systems, the original MVGPR is defined over a single input \cite{song2023modal}, time; for the general case of multi-dimensional inputs, see Refs. \cite{Sadagopan2021a, Sadagopan2022}.

\subsubsection{Univariate GPR}
To establish the terminology, we start with the relatively simple univariate version of GPR (UVGPR). The univariate GPR model is based on the following Gaussian process model \cite{Rasmussen2006,Forrester2008},
\begin{equation}
  g_{UV}\sim \GP(m(t),k(t,t'))
\end{equation}
where $m(t)$ is a scalar mean function that captures the global trend in the data and $k(t,t')$ is a scalar correlation function to account for the deviation of the data from the global trend.  A non-zero mean function may be useful for capturing transient dynamics for flows in non-equilibrium, but for statistically converged unsteady flows we assume $m(t)=0$ and $k(t,t')=k(|t-t'|)\equiv k(\tau)$. This assumption will be employed in the following.

Consider a dataset $\cD=\{(t^{(i)},\vg^{(i)})\}_{i=1}^{N_s}$ which can also be represented as $\cD = \{\vt, \vX\}$ where $\vt = [t^{(1)}, t^{(2)}, \cdots, t^{(N_s)}]^T$ and $\vX = [\vg^{(1)}, \vg^{(2)}, \cdots, \vg^{(N_s)}]^T$. The covariance of the system response can be calculated using the prescribed kernel function, $\text{cov}(\vg, \vg') = k(\vt, \vt^T)$.  One UVGPR $x_{UV,j}$ is generated for the $j$th output; and for a new input $t^*$ the prediction is given by the Gaussian distribution, $x_{UV,j}(t^*)\sim \cN(\mu_j(t^*;\cD),\sigma_j^2(t^*;\cD))$ where the mean and variance are represented as
\begin{equation}\begin{split}
    \mu_j(t^*; \cD) &= k(t^*, \vt) \left[k(\vt, \vt) + \sigma_n^2 \vI \right]^{-1} \vX_j \\ 
    \sigma_j(t^*; \cD) &= k(t^*, t^*) - k(t^*, \vt) \left[k(\vt, \vt) + \sigma_n^2 \vI \right]^{-1}k(\vt, t^*)
\end{split}
\end{equation}
where $\vX_j$ is $j$th column of $\vX$, and $\vI$ is $N_s \times N_s$ identity matrix.  The posterior distribution of the kernels parameters, $\vtb$ is proportional to the likelihood times the prior distribution, $p(\vtb|\; \vX, \vt) \propto p(\vX |\; \vt, \vtb) p(\vtb)$ where $p(\vX | \; \vt, \vtb)$ is likelihood of the model formed by data and kernel parameters, and $p(\vtb)$ is prior distribution of the kernel parameters. The training of the Gaussian process model can be achieved by minimizing the objective function, $\vtb^* = \argmin_{\vtb} \cL(\vtb)$. In this case, the objective function is negative log posterior distribution, $\cL(\vtb) = - \log p(\vtb|\; \vX, \vt)$.

\subsubsection{Multivariate GPR}\label{sf_mvgpr}
For cases where multiple outputs are involved, while it is possible to train one UVGPR for the prediction of each output, a more efficient approach is to employ a multivariable GPR (MVGPR) formulation.  The MVGPR also accounts for the correlation between the outputs and arguably provides higher accuracy than its univariate counterpart.  The MVGPR relies on a multivariate GP,
\begin{equation}
  \vg_{MV}\sim \GP(\mathbf{0},\vK(\tau))
\end{equation}
where $\mathbf{0}$ is a vector of zero entries, and $\vK(\tau)$ is a matrix of $N_y\times N_y$ correlation functions ($\tau = |t - t'|$ is used for stationary process).  The $(i,j)$th element of $\vK$ indicates the correlation between the $i$th and $j$th outputs.  Similar to the univariate case, the prediction at a new input $t^*$ is represented using a multivariate Gaussian distribution; see Refs. \cite{Parussini2017,Sadagopan2021a,Sadagopan2022} for details.

To guarantee the well-posedness of the MVGPR model, the correlation function needs to be chosen carefully.  One possibility is the linear model of coregionalization (LMC).  A rank-$N_K$ LMC model is a weighted sum of $N_K$ scalar correlation functions $\{k_j(\tau)\}_{j=1}^{N_K}$ \cite{Alvarez2012,Parussini2017},
\begin{equation}\label{eqn_lmc}
    \vK(\tau)=\tilde{\vtG}[\mathrm{diag}(k_1,k_2,\cdots,k_{N_K})]\tilde{\vtG}^T=\sum_{j=1}^{N_K}\tilde{\vtg}_j\tilde{\vtg}_j^T k_j(\tau)
\end{equation}
where the weight matrix $\tilde{\vtG}$ is of size $N_y\times N_K$, and its $j$th column vector $\tilde{\vtg}_j$ represents the correlation between the outputs and has to be estimated during the training of the MVGPR.  The $N_y$ scalar correlation functions can be different or partially same.  One can choose either $N_K<N_y$ to reduce the number of parameters in the matrix correlation function, or $N_K\geq N_y$ to capture complex correlations between the outputs.

The MVGPR with the LMC kernel in Eq. (\ref{eqn_lmc}) is equivalent to the linear combination of $N_K$ independent UVGPR's,
\begin{equation}\label{eqn_mv_uv}
  \vg_{MV}(t) = \tilde{\vtG}\vg_{UV}(t)
\end{equation}
where $g_{UV,i} \sim \GP(0,k_i(\tau))$.  This relation also shows that, if the correlation between the outputs is ignored, i.e., if $\tilde{\vtG}$ is a $N_y\times N_y$ identity matrix, the MVGPR reduces to $N_y$ UVGPR's, where the UVGPR models are generated for each of the outputs independently. It is also noteworthy that Eq. \eqref{eqn_mv_uv} also performs a dimensionality reduction of high-dimensional output data, where the columns of the weight matrix $\tilde{\vtG}$ are effectively the dominant modes of the data.  The MVGPR identifies the dominant modes and generates the mapping from input to modal coordinates \textit{simultaneously} \cite{Sadagopan2022}. As a result, the MVGPR modes are different from the modes obtained in standard methods, e.g., POD or principle component analysis (PCA), because the latter does not account for the distribution of inputs.

\section{Identification of Stationary Flow System via MVGPR}\label{sec3}

In this section, using the LMC kernel of MVGPR, we derive a system identification algorithm for the stationary flow dynamics.  To clarify, the stationary flow refers to the fluid flow whose statistical moments are time-invariant.

\subsection{Correlation function of stationary flow}

\subsubsection{1D simple harmonic oscillator}

We first derive the kernel function which corresponds to 1D oscillatory motion.  The correlation function of 1D oscillatory system can be derived from the 2D coupled harmonics system shown in Eq. \eqref{eqn_cho}

\begin{equation}\label{eqn_cho}
\left\{
    \begin{array}{l}
      \ddt{x_j(t)} = -j \omega_0 y_j(t) \\
      \ddt{y_j(t)} = j \omega_0 x_j(t)
    \end{array}
\right.
\end{equation}
where $\omega_0$ is characteristic frequency of the system, and the solution can be expressed as

\begin{equation}
\hat{\vg}_j(t) = 
\begin{bmatrix}
\cos(j\omega_0 t ) & -\sin(j\omega_0 t) \\
\sin(j\omega_0 t)  & \cos(j\omega_0 t)
\end{bmatrix} \hat{\vg}_j(0)
\end{equation}
where $\hat{\vg}_j = [x_j, y_j]^T, \; \hat{\vg}_j(0) \sim \cN(\mathbf{0}, \sigma_j^2\vI)$. The correlation function of $\hat{\vg}_j(t)$ is, by definition, $\bE_{\hat{\vg}_j(0)}(x_j(t) x_j(t + \tau))$, which will simply be denoted as $\bE(x_j(t) x_j(t + \tau))$ for convenience. Note that the distribution of initial condition is i.i.d, and hence the expectation value of any cross terms gives 0. 

\begin{equation}\begin{split}\label{eqn_cosine_der}
    \bE(x_j(t) x_j(t + \tau)) &= \bE(x_j(0) \cos(j \omega_0 t) - y_j(0) \sin(j \omega_0 t))(x_j(0) \cos(j \omega_0 (t+\tau)) - y_j(0) \sin(j \omega_0 (t+\tau))) \\
    &= \bE\{x_j(0)^2\cos^2(j\omega_0 t)\cos(j\omega_0 \tau) + y_j(0)^2\sin^2(j\omega_0 t)\cos(j\omega_0 \tau)\} \\
    &= \bE \{x_j(0)^2 \cos^2(j \omega_0 t) + y_j(0)^2 \sin^2(j \omega_0 t) \}\cos(j \omega_0 \tau) \\
    &= \sigma_j^2\cos(j\omega_0 \tau) 
    \end{split}
\end{equation}

Similarly, the correlation function of $y_j(t)$ is $\sigma_j^2\cos(j\omega_0 \tau)$. Also, the correlation between the different states can be derived by $\bE(x_j(t)y_j(t+\tau))$ and gives ${\sigma_j^2\sin(j\omega_0\tau)}$, which explains that the harmonics conjugate has to have a phase difference by $90\degree$. In contrast, $\bE(x_j(t+\tau)y_j(t))$ gives $ -{q_j^2\sin(j\omega_0\tau)}$. For those who are interested in more theoretical backgrounds of periodic correlation function, see Ref. \cite{solin2014explicit}.

\subsubsection{Modal Decomposition of Multi-dimensional Linear Dynamics}

Next, under Koopman invariant subspace, a finite dimensional Koopman generator $\mathbb{K}$ maps from observable $\vg \in \bR^{n}$ to its time derivative $\dot{\vg}$,
\begin{equation}
    \dot{\vg} = \mathbb{K} \vg.
\end{equation}

Assuming there exists an eigendecomposition of $\mathbb{K}$,
\begin{equation}
    \dot{\vg} = \vtG \vtL \vtY^H \vg  
\end{equation}
where $\vtG = [\vtg_1, \overline{\vtg}_1,\; ...,\; \vtg_{N_k}, \overline{\vtg}_{N_k}]$ is right eigenvector matrix, $\vtY = [\vty_1, \overline{\vty}_1, ..., \vty_{N_k}, \overline{\vty}_{N_k}]$ is left eigenvector matrix, and $\vtL$ is eigenvalue matrix of the Koopman operator. In this paper, we restrict ourselves into stationary flow, so all the eigenvalues in $\vtL$ are imaginary values. The line over the symbol stands for complex conjugate. Then, the solution trajectory can be formed as
\begin{equation}\label{ed}
    \vg_t = \vtG \vD_t \vtY^H \vg_0
\end{equation}
where $\vD_t = \exp(\vtL t)$ and $\vg_t$ is observable state after $t$ times from $\vg_0$. This can be reformulated in terms of real-valued matrices.

\begin{equation}
    \vg_t = \tilde{\vtG}\tilde{\vD}_t \tilde{\vtY}^T \vg_0
\end{equation}
where $\tilde{\vtG} = \sqrt{2} [\vtg_{1, R}, \vtg_{1, I},\; ... , \vtg_{N_k, R}, \vtg_{N_k, I}]$, $\tilde{\vtY} = \sqrt{2} [\vty_{1, R}, \vty_{1, I}, ..., \vty_{N_k, R}, \vty_{N_k, I}]$ and the second subscript $R$ and $I$ stand for real and imaginary part respectively. The matrix $\tilde{\vD}_t$ is a block diagonal matrix where each block represents each independent harmonic process as shown in Eq. \eqref{ind_harmonics}.

\begin{equation}\label{ind_harmonics}
    \tilde{\vD}_t = \begin{bmatrix}
        \cos(\omega_1 t) & -\sin(\omega_1 t) & & & & \\
        \sin(\omega_1 t) & \cos(\omega_1 t) & & & &\\
        & & \ddots & &\\
        & & & \cos(\omega_{N_k}t) & -\sin(\omega_{N_k} t) \\
        & & & \sin(\omega_{N_k} t) & \cos(\omega_{N_k} t)
    \end{bmatrix}
\end{equation}

Consider the real-valued modal coordinates, $\vtf$ which is the observable state mapped onto $\tilde{\vtY}$,
\begin{equation}\label{eq:modal_coordinate}
    \vtf = \tilde{\vtY}^T \vg_0 = [\phi_1, \phi_2, \; ... ,\phi_{2N_k}]^{T}.
\end{equation}

The elements in $\vtf$ are treated as i.i.d zero mean Gaussian random variables, $\phi_{2k-1}, \phi_{2k} \sim \cN(0, \sigma_k^2)$ and $k=1,2,...,N_k$ so that the same statistical approach used in the 1D case can be applied.

\subsubsection{Correlation function of stationary flow system}

Subsequently, consider an observable, $\vg_0(\vtx)$ defined on a spatial coordinate $\vtx \in \Omega$. The correlation function at two different instance of time, $t$ and $t + \tau$ at two different locations in $\Omega$, $\vtx_i$ and $\vtx_j$ is denoted as $\vC(t, t+\tau, \vtx_i, \vtx_j)$. This function can be discretized so that $\vC(t, t+\tau, \vtx_i, \vtx_j) = \vC_{i,j}(t, t+\tau)$, and by definition it can be represented as

\begin{equation}\label{eqn_exp_corr}
    \vC(t, t + \tau) = \bE\left[\vg_t \vg^T_{t+\tau}\right] = \tilde{\vtG}\bE\left[(\tilde{\vD}_t\vtf)(\tilde{\vD}_{t+\tau}\vtf)^T\right]\tilde{\vtG}^T
\end{equation}
where $\vtf = \vtf(\vg_0)$. Note that the product $\tilde{\vD}_t \vtf \in \bR^{2N_k}$ is a vector that contains $\sin(\omega_k t)$ and $\cos(\omega_k t)$ with coefficients of $\phi_{j, R}$ and $\phi_{j, I}$. The product of $\tilde{\vD}_t\vtf$ and $\tilde{\vD}_{t+\tau}\vtf$ produces a similar derivation to the 1D case, which reduces the dependence on two temporal points down to the temporal difference between the points, $\vC(t, t+\tau)=\vC(\tau)$.

\begin{equation}\label{eqn_mperker}
    \vC(\tau) = \tilde{\vtG} \vM \tilde{\vtG}^T, \quad \vM =
    \mathrm{diag}[\sigma_1^2\cos(\omega_1 \tau), \sigma_1^2\cos(\omega_1 \tau), \hdots, \sigma_{N_k}^2\cos(\omega_{N_k} \tau))]
\end{equation}

It should be noted that ideally there exists the cross-correlation between the complex conjugate states which can potentially make $\vM$ a block-diagonal matrix. However, regardless of the cross-correlation, the identified modes will be equivalent because one is the result of the unitary rotation of the other one in complex domain, and hence the resulting modes span the same subspace.

In fact, the MVGPR model described in Sec. \ref{sf_mvgpr} can only work with a single realization. To be specific, the model does not have a capability of differentiating the data from different realizations, and knowing the temporal difference between the data points is not sufficient to inform their phase to the model. This can be critical for modal analysis purpose when multiple realizations need to be considered to increase the model accuracy. To address this issue, more advanced model is derived as a result of system identification of MVGPR.

\subsection{Approximate MVGPR model via LMC kernel}\label{aMVGPR_LMC}

The LMC kernel has a similar structure of the correlation function derived from multi-variate harmonics system. In order to account for the multiple trajectories using MVGPR model, the LMC kernel needs to incorporate the phase information of different trajectories. This can be achieved by introducing an appropriate mapping between observable and latent states. One way to tackle this issue is to make use of linear kernel that takes the observable state, $\vg$, and is parameterized by corresponding weights, $\vw_j$. Using this form of kernel function, we are inducing a bias in the model such that the coordinates after mapping onto the weight $\vw_j$ is the modal coordinates defined in Eq. \eqref{eq:modal_coordinate}. Given a pair of realizations of observable denoted as $\vg_0'$ and $\vg_0''$, one can construct a linear kernel.

\begin{equation}
    k_{\mathrm{lin, j}}(\vg_0', \vg_0'') = (\vw_j^T \vg_0')(\vw_j^T \vg_0'')
\end{equation}

The new LMC kernel contains $k_j(\tau)k_{\mathrm{lin}, j}(\vg_0', \vg_0'')$ in its diagonal entries where $k_j$ is usually a cosine kernel for a stationary flow with a particular frequency component, $\omega_j$. The same linear kernel needs to be used for conjugates, e.g. $k_{\text{lin}, 1}=k_{\text{lin}, 2}$, $k_{\text{lin}, 3}=k_{\text{lin}, 4}$ and so on. As a result, we have a new form of LMC kernel,
\begin{equation}\label{eqn_newlmc}
    \vK_{\mathrm{LMC}}(\tau, \vg_0', \vg_0'') =  \sum_{j=1}^{2N_k}  \tilde{\vtg}_j \left[ k_j(\tau) k_{lin,j}(\vg_0', \vg_0'')  \right] \tilde{\vtg}_j^T
\end{equation}

In this way of representation, one can take the stochastic average of the kernel function to recover the correlation function of the system. In Eq. \eqref{cor_stc_avg}, $\vK_{\mathrm{LMC}}$ represents the correlation of a particular realization while $\vC$ is the averaged correlation of the system. Consider the expectation of $\vK_{\mathrm{LMC}}$ over $\vtf$ \cite{edition2002probability},

\begin{equation}\begin{split}\label{cor_stc_avg}
    \bE\left[\vK_{\mathrm{LMC}}(\tau, \vg_0, \vg_0)\right] &= \sum_{j=1}^{2N_k}  \tilde{\vtg}_j \left[ k_j(\tau) \int_{\bR} \int_{\bR}  \phi_j' \phi_j'' f_{\phi_j}(\phi_j', \phi_j'')  d\phi_j' d\phi_j'' \right] \tilde{\vtg}_j^T \\
    &= \sum_{j=1}^{2N_k} \sigma_j^2 k_j(\tau) \tilde{\vtg}_j \tilde{\vtg}_j^T \\
    &= \vC (\tau)
\end{split}
\end{equation}
where $f_{\phi_j}(\phi_j', \phi_j'')$ is a joint probability density of $\phi_j' = \vw_j^T \vg_0'$ and $\phi_j'' = \vw_j^T \vg_0''$ which is Gaussian with variance of $\sigma_j^2$. Note that the expectation over the random variables was already carried out in the derivation of LMC kernel, Eq. \eqref{eqn_exp_corr}. However, by introducing the mapping through the linear kernel, it is possible to preserve only the necessary part of the correlation. The possible confusion that may arise is that even though $\vK_{\mathrm{LMC}}$ is the correlation for one particular realization of the system, it does not mean that the phase has to be treated as a deterministic quantity, but instead such phase will be determined by $\vw_j$ given the measurement pair \{($\vg_0'$, $\vg_0''$)\}. As a consequence, the averaged correlation of the system will be recovered when the random variables are marginalized out as shown in Eq. \eqref{cor_stc_avg}.

\begin{figure}[H]
\centering
\includegraphics[width=\textwidth]{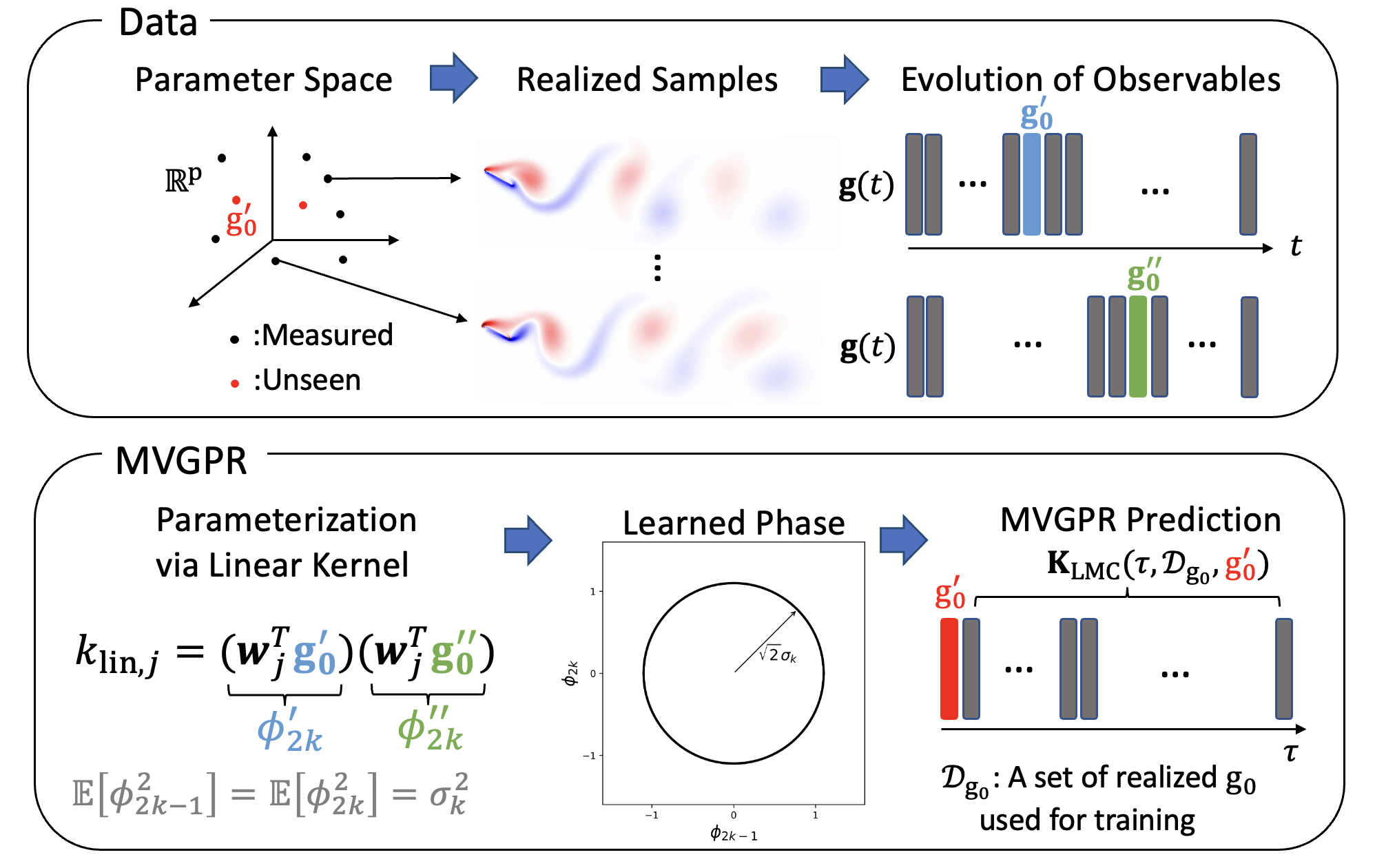}
\caption{Schematic of MVGPR model with linear kernel parameterization}\label{fig:lin_ker}
\end{figure}

Figure \ref{fig:lin_ker} illustrates the role of linear kernel, parameterization of observables and mapping to modal coordinates. The samples are realized within the parameter space in $\bR^p$; the example of the parameter here is the phase angle between pitch and plunge motion of an airfoil in flow.  The prediction of MVGPR can be done by identification of phase via linear kernel.

\subsection{SPOD and MVGPR}
To conceptually establish the connection between SPOD and MVGPR, it is helpful to recall that both methods rely on the correlation of the data. In SPOD method, the correlation tensor is constructed empirically by taking the average of different flow realizations while in MVGPR the prescribed kernel functions are used to capture such correlations. However, both contain consistent information because the the correlation function defined in SPOD is the stochastic average over all the correlation of the flow realizations. The fundamental difference between the two methods can be viewed in frequentist vs Bayesian perspective, two competing philosophies in statistics. In frequentist's point of view, the kernel parameters can be estimated by a number of trials and observe how likely the event occurs without prior knowledge. SPOD conducts Fourier transform of multiple sets of data and considers the average of their spectral densities. This then measures how likely the data contains certain frequency component, which is essentially what SPOD eigenvalues represent in the spectral domain. Whereas the MVGPR model relies on Bayes' theorem to compute the posterior distribution of the the parameters using the prior belief of their distributions. Therefore, it is classified as a Bayesian inference method. In the following, we will elaborate the theoretical connection between the two methods.

\subsubsection{SPOD formulation}
When the flow is wide-sense stationary, which is usually the case in many turbulent flows \cite{arbabi2017study}, one can decompose the flow data in both space and time domain. Unlike space-only POD in which the inner product is defined only in the spatial domain, in spectral POD following space-time inner product is defined as

\begin{equation}
    \langle \vq, \vp \rangle_{\vtx, t} = \int_{-\infty}^{\infty} \int_{\Omega}{\vp^H(t, \vtx)\vq(t, \vtx)} d\vtx dt
\end{equation}
where $\vtx \in \Omega$ is a spatial coordinate which defines the flow field. Then spectral POD obtains the set of basis functions that maximize the following quantity 

\begin{equation}
    \lambda = \max_{\vtvf}\frac{\bE[|\langle \vq(t, \vtx), \vtvf(t, \vtx)\rangle|_{\vtx, t}^2]}{\langle \vtvf(t, \vtx), \vtvf(t, \vtx)\rangle_{\vtx, t}}
\end{equation}
where $\vq(t, \vtx)$ is a flow of interest defined in $\Omega$. The expectation is taken over the probability space in which different stochastic realizations exist. The new optimal space is formed by the set of eigenfunctions, $\vtvf_j(t, \vtx)$, and they are orthogonal with respect to the space-time domain. The eigenfunction can be solved by considering an appropriate eigenvalue problem,

\begin{equation}
    \int_{-\infty}^{\infty}\int_{\Omega}\vC(t, t', \vtx, \vtx') \vtvf(t', \vtx') d\vtx' dt' = \lambda \vtvf(t, \vtx).
\end{equation}
where the correlation tensor is represented as
\begin{equation}
    \vC(t, t', \vtx, \vtx') = \bE[\vq(t, \vtx) \vq^H(t', \vtx')].
\end{equation}

For stationary flow, due to its transnational invariance, the dependence of $t$ and $t'$ reduces to the difference between them, $\vC(t, t', \vtx, \vtx') = \vC(\tau, \vtx, \vtx')$ where $\tau = |t - t'|$. As their statistical property is invariant under translation, the integral of the correlation is not well defined, $\int_{-\infty}^{\infty}\int_{\Omega}{\vC(\tau, \vtx, \vtx')}d\vtx'd\tau \nless \infty$. Nevertheless, the spectral representation of the correlation is indeed well defined and hence can be used to find the eigenfunctions in the spectral domain. According to the Wiener-Khinchin theorem, the corresponding spectral density function can be derived by the Fourier transform of the correlation function.
\begin{equation}\label{eqn:csd}
    \vS(f, \vtx, \vtx') = \int_{-\infty}^{\infty}{\vC(\tau, \vtx, \vtx')} e^{-2\pi if\tau}d\tau
\end{equation}

Then, the SPOD modes are the eigenfunctions of the following eigenvalue problem,

\begin{equation}\label{eqn:spec_evp}
    \int_{\Omega}{\vS(f, \vtx, \vtx')\vu(f, \vtx')} d\vtx' = \lambda(f)\vu(f, \vtx)
\end{equation}

The optimal expansion of the flow in spectral domain can be represented as following,
\begin{equation}
    \hat{\vq}(f, \vtx) = \sum_{j=1}^{\infty}a_j(f)\vu_j(f, \vtx)
\end{equation}
where $a_j(f) = \langle \hat{\vq}(f, \vtx), \vu_j(f, \vtx)\rangle_{\vtx}$. The spectral density function can be represented by superposition of infinite number of eigenfunctions.
\begin{equation}\label{eqn:spec_density}
    \vS(f, \vtx, \vtx') = \sum_{j=1}^{\infty}\lambda_j \vu_{j}(f, \vtx)\vu^H_{j}(f, \vtx')
\end{equation}
where the eigenvalue can be represented as

\begin{equation}
    \bE[a_j(f) a_k^*(f')] = \lambda_j \delta_{jk} \delta(f - f')
\end{equation}

The rank of the orthonormal SPOD modes are assigned based on the eigenvalues.

\subsubsection{Relationship between SPOD and MVGPR}
As discussed previously, even though both SPOD and MVGPR rely on correlation function, their correlation functions are defined differently. In SPOD, the correlation is computed by averaging the ensemble stochastic flow data while in MVGPR, such statistics is induced by the prescribed kernel functions. In order to mathematically show the connection between MVGPR and SPOD, we need to revisit the spectral density function from its definition which is the Fourier transform of the correlation function,

\begin{equation}
    \vC(\tau) = \bE\left[\vg_{t} \vg_{t+\tau}^H\right] = {\bE\left[\vtG \exp(i \Omega t)\vtf \vtf^H \exp(-i \Omega (t+\tau))\vtG^H \right]}
\end{equation}
where $\Omega = \mathrm{diag}[\omega_1, -\omega_1, ..., -\omega_{N_k}]$, and $i$ is imaginary number. The expectation is taken over the random vector $\vtf$ which will provide following representation.

\begin{equation}
    \vC = \vtG \begin{bmatrix}
        \sigma_1^2 \exp(i \omega_1 \tau) & &  \\
        & \sigma_1^2 \exp(-i \omega_1 \tau) & &  \\
        & & \ddots &  \\ 
        & & & \sigma_{N_k}^2 \exp(-i \omega_{N_k} \tau) \\
    \end{bmatrix} \vtG^*
\end{equation}

Since the Fourier transform of $\exp(i\omega_j \tau)$ and $\exp(-i\omega_j \tau)$ are $2\pi \delta(\omega - \omega_j)$ and $2\pi \delta(\omega + \omega_j)$ respectively, the spectral density function will be

\begin{equation}\label{spod_spd}
    \vS = \vtG \vtD\vtG^H, \quad \vtD = \mathrm{diag}[2\pi\sigma_1^2\delta(\omega - \omega_1), 2\pi\sigma_1^2 \delta(\omega + \omega_1), ..., 2\pi\sigma_{N_k}^2 \delta(\omega - \omega_{N_k})]
\end{equation}


Now, this formulation can be generalized for the case of multiple modes per frequency. Consider a spectral density function which contains $N_m$ modes at frequency $\omega_j$.

\begin{equation}\label{spec_gp}
\vS(\omega_j, \vtx, \vtx') = \vtG_{j} \begin{bmatrix}
        2\pi \sigma_1^2& &  \\
        & 2\pi \sigma_1^2 & &  \\
        & & \ddots &  \\ 
        & & & 2\pi \sigma_{N_m}^2 \\
    \end{bmatrix} \vtG_{j}^H
\end{equation}
where $\vtG_j$ is a set of LMC modes at $\omega_j$. Note that the spectral density function is a normal matrix ($\vS\vS^H = \vS^H\vS$). This property provides a unique rank $N_m$ unitary decomposition where the matrix in the middle has non-zero values only in the diagonal components as shown in Eq. \eqref{unit_decomp}

\begin{equation}\label{unit_decomp}
    \vS(\omega_j, \vtx, \vtx') = \mathcal{U}_j\underline{\Lambda}_j \mathcal{U}_j^H
\end{equation}
where $\mathcal{U}_j$ is a set of $N_m$ orthonormal vectors or eigenvectors associated with $\omega_j$, and $\underline{\Lambda}_j = \mathrm{diag}(\underline{\lambda}_{j, 1}, \underline{\lambda}_{j, 1}, \hdots, \underline{\lambda}_{j, N_m})$ contains corresponding eigenvalues. Here, \eqref{spec_gp} and \eqref{unit_decomp} are the same spectral density function with different basis vectors. Therefore, one can verify that there exists a linear transformation, $\vT$ such that 

\begin{equation}
    \vtG_j = \mathcal{U}_j\vT
\end{equation}
where $\vT$ is a $N_m \times \N_m$ matrix with full rank. Therefore, both $\mathcal{U}_j$ and $\vtG_j$ span the same subspace. In general, one can find $\vT$ for any $\omega_j$. Therefore, this shows that the identified LMC weights and SPOD eigenvectors span the same subspace for every $\omega_j$.

\subsubsection{MVGPR modes as SPOD modes}

In fact, the MVGPR model assumes nothing about the target modes while the SPOD model captures statistically optimal modes associated with each frequency component. Therefore, when MVGPR model is trained, it tends to identify any modes that span the same subspace, which may lead to loss of dynamically important features of the system such as phase information. To tackle this challenge, we augment orthonormal regularization term to the objective function so that the model is forced to find the eigenvectors. The new objective function is represented as

\begin{equation}\begin{split}
    \vtb^* = \argmin_{\vtb}& \cL(\vtb) + \lambda_{\cR} \cR(\vtG) \\
    \cR(\vtG) &= ||\vtG^H \vtG||_{F}^2
\end{split}
\end{equation}
where $||.||_F$ denotes Frobenius norm, and $\lambda_{\cR}$ is a regularization factor. Currently, the MVGPR model does not work with complex values, but $\cR(\vtG)$ can be constructed by real-valued matrices, $\vtG = \tilde{\vtG}_R + j \tilde{\vtG}_I$.

\begin{equation}\begin{split}
\vtG^H\vtG &= \tilde{\vtG}_R^T \tilde{\vtG}_R + \tilde{\vtG}_I^T \tilde{\vtG}_I  + j \left(\tilde{\vtG}_R^T\tilde{\vtG}_I - \tilde{\vtG}_I^T \tilde{\vtG}_R \right) \\
||\tilde{\vtG}^H\tilde{\vtG}||_F^2 &= ||\tilde{\vtG}_R^T \tilde{\vtG}_R + \tilde{\vtG}_I^T \tilde{\vtG}_I - \vI||_F^2 + ||\tilde{\vtG}_R^T\tilde{\vtG}_I - \tilde{\vtG}_I^T \tilde{\vtG}_R ||_F^2
\end{split}
\end{equation}

Finally, the rank of the MVGPR modes can be calculated based on the identified variances and the weights of the linear kernel. Assuming $\vg_0$ is i.i.d and coordinate-wise normalized, $||\vw^T_j\vg_0||_2^2 = ||\vw_j||_2^2$ where $||.||_2$ denotes L2 norm; this will be true if $\vg_0$ is obtained from standard POD with appropriate normalization. Utilizing this, one can find the eigenvalues corresponding to the MVGPR modes.

\begin{equation}\begin{split}\label{mvgpr_eig}
    k_j(\tau, \vg_0', \vg_0'') &= \sigma_j^2 \cos(\omega_j \tau) (\vw_j^T \vg_0') (\vw_j^T \vg_0'') \\
    \Rightarrow \tilde{\lambda}_j &=  \sigma_j^2 ||\vw_j||_2^2
\end{split}
\end{equation}

The calculated eigenvalues, $\tilde{\lambda}_j$ may not be exactly equivalent to $\lambda_j$ from SPOD, but it still provides similar mathematical quantity. Therefore, if the model is trained properly, the weights of the linear kernel and the variance should be able to give the eigenvalues.

\subsection{MVGPR for Modal Analysis}
Hitherto, the connections among MVGPR, linear systems, Koopman analysis, and SPOD have been established. Basically, training a MVGPR model with an appropriate LMC kernel on a dataset of trajectories is shown to produce a set of moedes that span the same subspace with the Koopman modes, and equivalently SPOD modes. Unlike the original version of MVGPR explained in Sec \ref{sf_mvgpr}, in the current version of MVGPR, the input is not only time but also the measurement data.

\begin{equation}
  \vx_{MV}\sim \GP(\mathbf{0},\vK(\tau, \vg_0', \vg_0''))
\end{equation}

We now provide a process of modal analysis using MVGPR. Consider a dataset of trajectories with $s$ realizations, $\cD = \{\cD_1, \cD_2, \hdots, \cD_s\}$ where $\cD_j=\{(t_i,\vg_i)\}_{i=0}^{m-1}$ is the $j$th realization of the system, and $m$ is the number of measurements in that realization. As illustrated earlier, MVGPR has two types of input: the measurement, $\vg_i$ and the time at which the measurement is collected, $t_i$. The two time instances are used to calculate the temporal difference, $\tau = |t_i-t_k|$, and hence there are actually three input to the MVGPR model, $\vg_i$, $\vg_k$, and $\tau$. During the training process, the cosine kernels learn the correlation between the temporal data points $(t_i, t_k) \in \cD$ while the linear kernels learn the correlation between the two measurements, $(\vg_i, \vg_k) \in \cD$, and they may belong to same realization or different realizations. In general, it is required to have sufficiently long temporal steps for each $\vg_i$ and $\vg_k$ so that adequate degree of information is provided for both kernels.  

Now consider another dataset, $\cD^* = \{(t_i^*, \vg_i^*)\}_{i=0}^{m-1}$ which is not a subset of $\cD$.  For a measurement point $\vg_0^*$, the trained MVGPR model constructs a mapping so that GP takes input of $\{\cD, (t_j^*,\vg_i^*)\}$ and returns $\vg^*_{i+j}$, which is the estimated value of observable at $t_j^*$ time future from $\vg_i^*$.
The algorithm for modal analysis based on MVGPR is formalized as follows,
\begin{enumerate}
    \item If the $\vg$'s are of very high dimension, it would be beneficial to project the data onto a reduced space, producing a new dataset $\hat{\cD}_j=\{(t_i,\hat{\vg}_i)\}_{i=0}^{m-1}$ where $\hat{\vg}=\vP\vg$.  The projector $\vP$ can be found by, e.g., standard POD procedure. Also, coordinate-wise normalization of $\hat{\vg}$ is usually preferred for better numerical behavior. In this case, $\vP$ needs to be scaled accordingly.
    \item The MVGPR requires data pre-processing such that if the input of the GP is $(t_k, \hat{\vg}_i) \in \hat{\cD}_j$, the corresponding output is $ \hat{\vg}_{i+k} \in \hat{\cD}_j$.
    \item Determine the number of modes of interest, and select the kernel for each LMC mode. For stationary flow, one can choose the cosine kernel Eq. (\ref{eqn_cosine_der}) and let the learning process decide the kernel parameters.
    \item Learn the MVGPR from data using a GP implementation, and the LMC kernel produces weight matrix $\tilde{\vtG}$ and frequency of the Cosine kernels.
    \item The Koopman modes or SPOD modes are recovered by $\vP^+\tilde{\vtG}$, and the Koopman eigenvalues are recovered by the identified frequencies of the Cosine kernels. The rank (weight) of the modes, $\tilde{\lambda}_j$ can be computed using Eq. \ref{mvgpr_eig}.
\end{enumerate}

The implementation of MVGPR code is based on the third-party Python package, GPflow.


\section{Results and Discussion}\label{sec4}
This section demonstrates the modal analysis methodology using MVGPR through a range of examples, including (1) the standard cylinder vortex flow problem, (2) a linear synthesized flow data having multiple modes per frequency, and (3) flow over a flat plate in both pitching and plunging motion.

\subsection{Problem 1: Cylinder Vortex Flow}
The first case problem is the cylinder flow field in limit cycle oscillation, that is publicly available \cite{kutz2016dynamic}. The spatial dimensions of data are $199 \times 449$, and there are 151 time steps with the data sampling frequency of 1/3Hz, which gives uniform time steps from 0 to 5 seconds. For the subsequent analyses, first 130 data points are used for training both MVGPR and DMD models. The orthonormal regularization factor, $\lambda_{\cR}$ is set to zero as there exists only single conjugate pair for each frequency component.

\subsubsection{Full Measurement Data}

\begin{figure}[H]
\centering
\includegraphics[width=\textwidth]{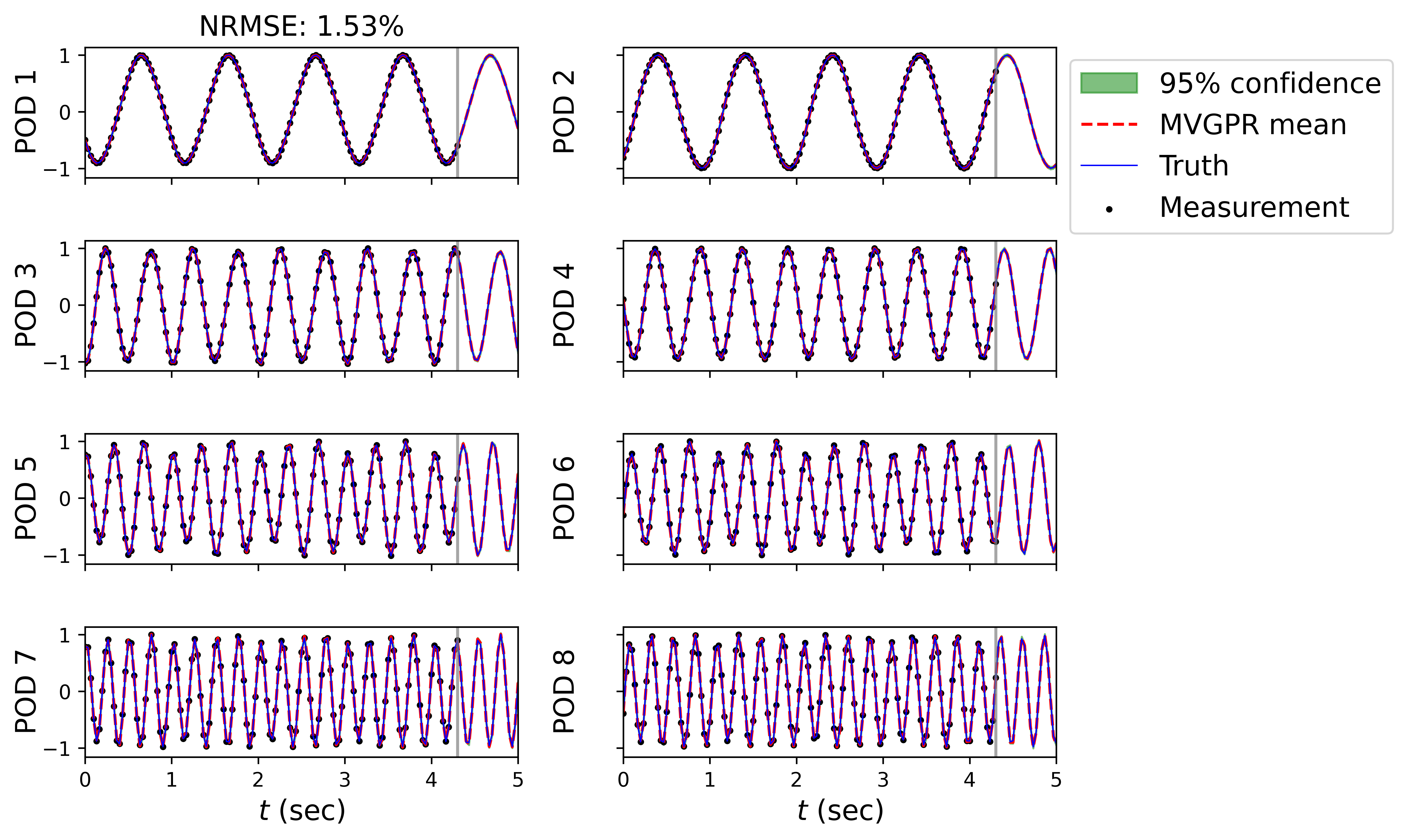}\hspace*{-2.5cm}
\caption{MVGPR mean and 95\% condidence interval of POD coordinates}\label{fig:cyn_gp_temporal}
\end{figure}

\begin{equation}
    \text{NRMSE} = \frac{|| \vg_{\text{true}} - \vg_{\text{est}}||_2}{||\vg_{\text{true}}||_2} \times 100 \%
\end{equation}

In order to visually show the correspondence between DMD and MVGPR modes, the identified mode shapes for each method are presented in Fig. \ref{fig:cyn_dmd_gp_modes}. In this analysis, 8 POD modes are used to capture approximately 99.6\% of the energy (training dataset). The DMD modes are complex-valued modes which involve complex-conjugate pair for each frequency component while in MVGPR the identified modes are real-valued Therefore, the two real-valued LMC weights are required to capture the real and imaginary component of the DMD modes. In Fig. \ref{fig:cyn_dmd_gp_modes}, the four complex-valued DMD modes are shown in order of increasing frequency (a), and the eight real-valued MVGPR modes are shown in the same way. One can clearly see that the real parts of DMD modes correspond to the odd numbered MVGPR modes, and the imaginary parts of DMD modes correspond to the even numbered MVGPR modes. 

\begin{figure}[H]
\centering
\insertfigs{cylinder_clean_dmd_modes_8}{0.47}{Four complex-valued DMD modes}
\insertfigs{cylinder_clean_gp_modes_8}{0.47}{Eight real-valued MVGPR modes}
\caption{Correspondence of DMD and MVGPR modes}
\label{fig:cyn_dmd_gp_modes}
\end{figure}

\begin{figure}[H]
\centering
\includegraphics[width=.7\textwidth]{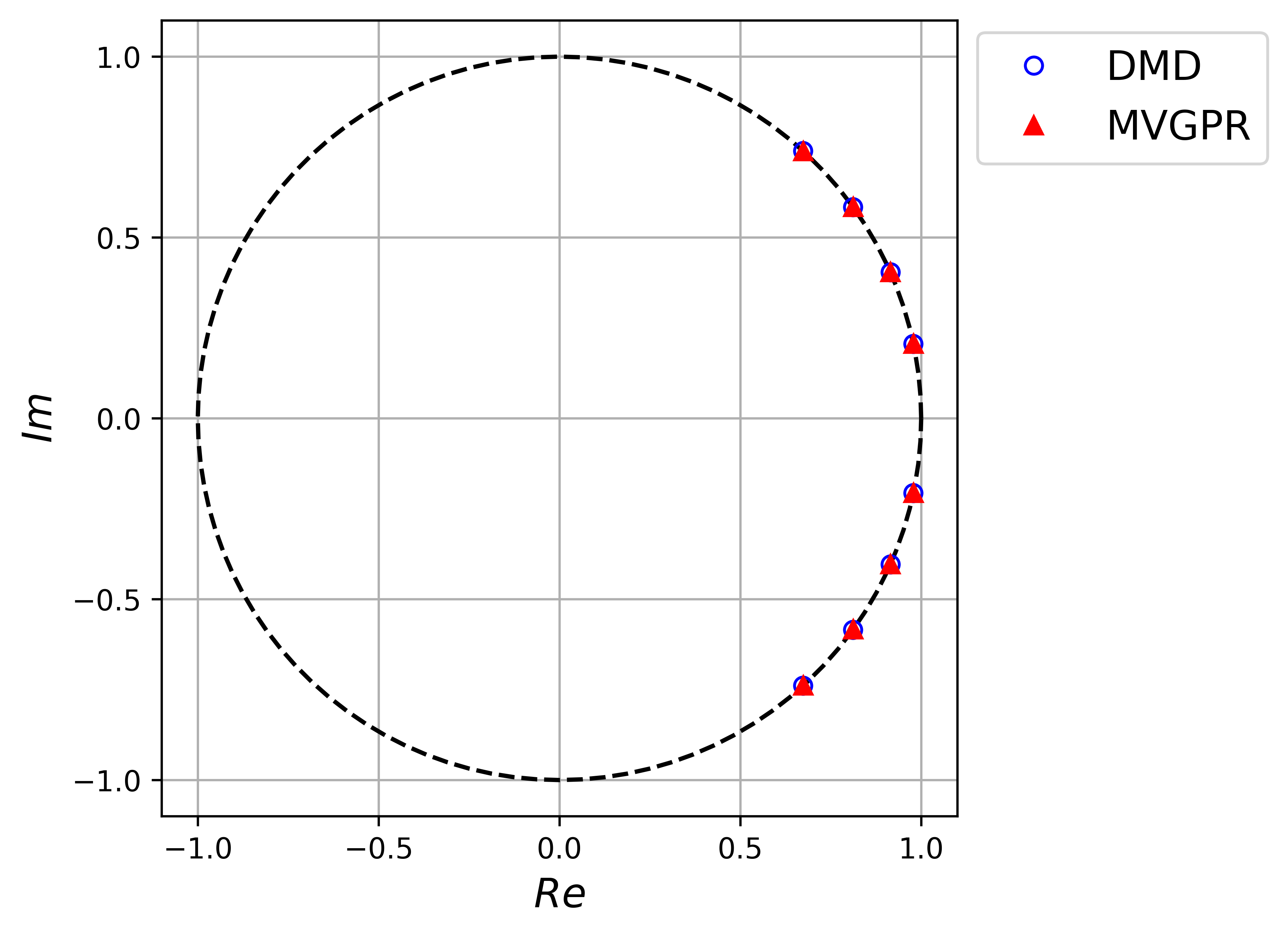}\hspace*{-2cm}
\caption{The identified frequencies of DMD and MVGPR}\label{fig:cyn_clean_eigs}
\end{figure}

\subsubsection{Temporally Irregular Measurement}

\begin{figure}[H]
\centering
\includegraphics[width=.8\textwidth]{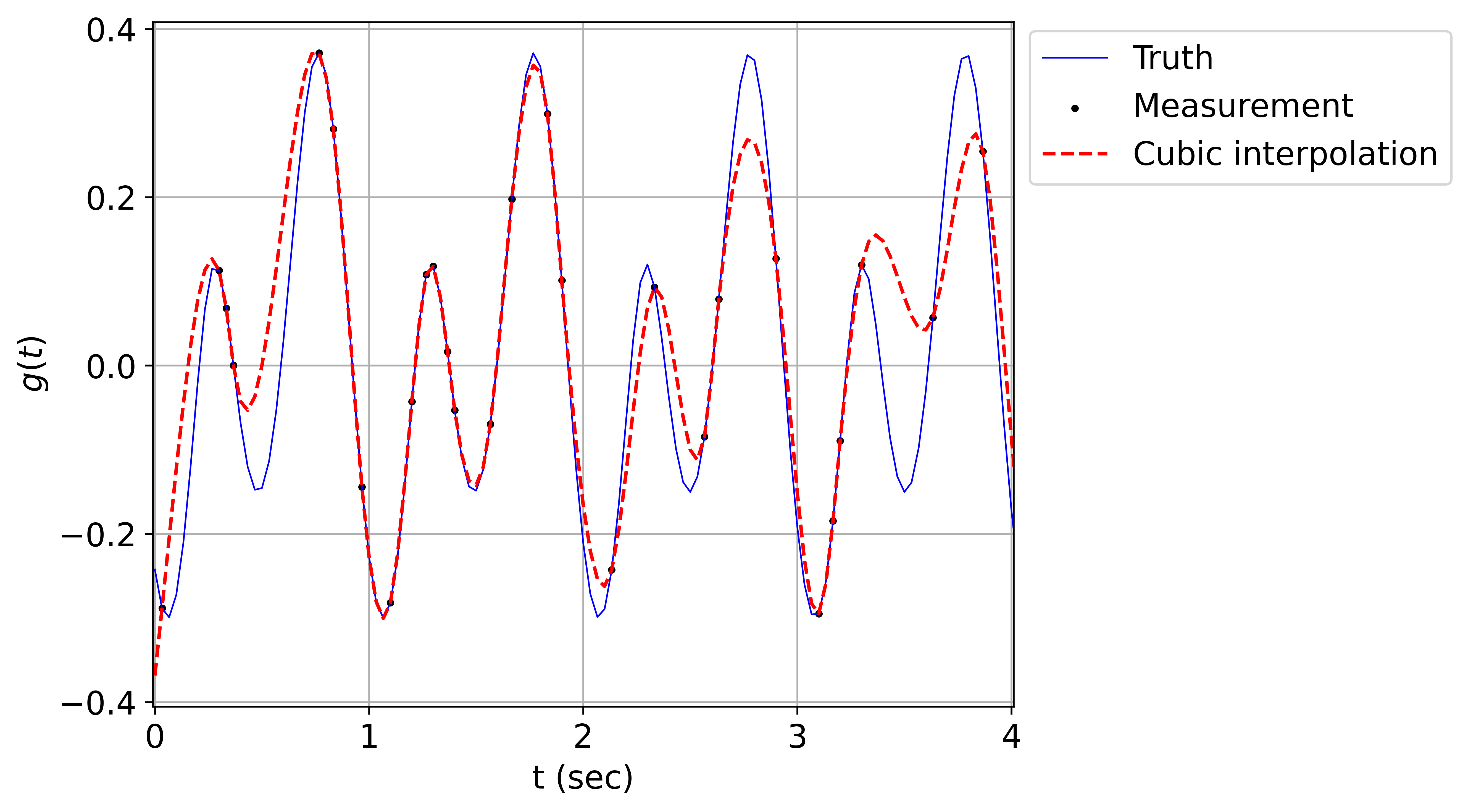}\hspace*{-2.5cm}
\caption{Reconstructed flow with cubic interpolation for 24\% measurement}\label{fig:cylinder_irr_measurement}
\end{figure}

This test case considers a challenging scenario especially for standard DMD, where the dataset is temporally irregular, i.e., the snapshots are non-uniformly distributed in time domain. The sparsity is achieved by randomly removing some of the data points from the original dataset. The percentages of retained data range from 50\% to 20\%. To see the effect of cubic interpolation, flow field at a particular location is shown in Fig. \ref{fig:cylinder_irr_measurement}. The full trajectory shown in red is calculated based on the measurement data points marked with black dots. One can clearly observe the difference between the true trajectory and interpolated one. This difference is expected to contaminate the spectral information of the system and hence worsen the identified modes. It should be noted that for this problem, the first version of the MVGPR model is employed.

To reduce the dimension of the problem, the data is projected onto a subspace of 8 POD modes. For the MVGPR model, 8 cosine kernels are employed where each pair contains the identical kernels to capture the harmonics. In DMD algorithm, it is not possible to directly apply it to this type of data. Therefore, cubic interpolation is used to reconstruct the dynamics on a uniform time grid, so that the conventional DMD algorithm can still be applied. The MVGPR and interpolated DMD results are compared with the true DMD modes from the original training dataset.

\begin{figure}[H]
\centering
\includegraphics[width=\textwidth]{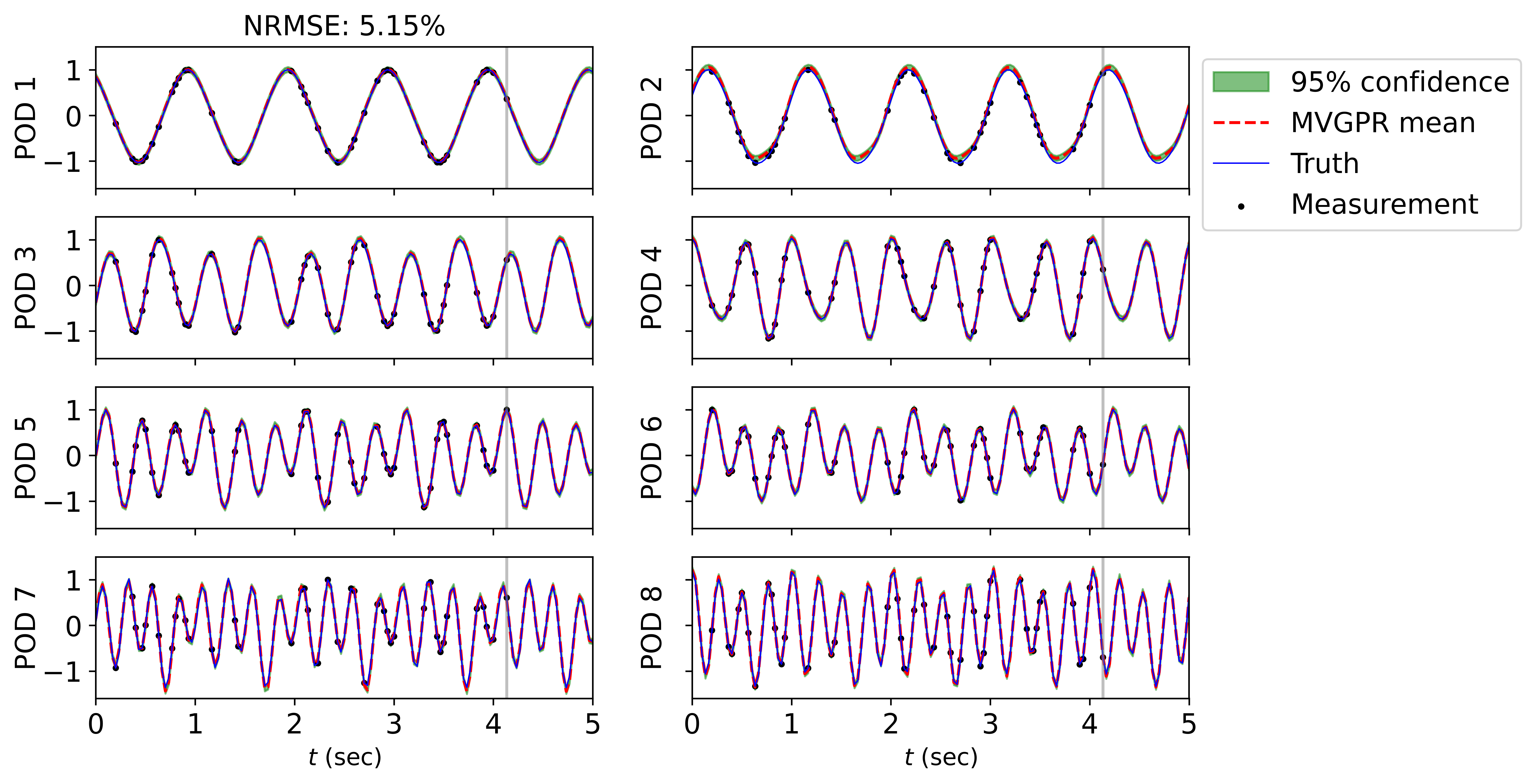}\hspace*{-2cm}
\caption{Training measurements and MVGPR mean and 95\% confidence interval}\label{fig:cylinder_irr_temp}
\end{figure}

First, the temporal prediction of the MVGPR model is examined for the case of 32\% data in Fig. \ref{fig:cylinder_irr_temp}, where the projected coordinates of the irregular data points are compared with the MVGPR mean prediction and 95\% confidence interval.  The first thing to note is that, due to the lack of data points, the projected coordinates now contain multiple frequency components. The MVGPR model accuracy is decreased compared to the full measurement case discussed earlier, but it is still able to capture the temporal dynamics with an confidence interval that is nearly zero over all time.  The near-zero interval is explained by the fact that GP assumes a periodic kernel structure and only needs to learn the information of one period from data; in the irregular dataset, while each period only contains a partial set of data points, the combination of data points from different periods is likely to densely cover the one entire period for the learning of the periodic kernels, and thus strengthen the confidence.

Subsequently, the mode shapes from MVGPR and interpolated DMD are compared in Fig. \ref{fig:dmd_gp_modes_sp}.  Note that post-processing of the mode shapes has been done using Grassmannian subspace for a meaningful comparison between them.  Even with only 20\% data, MVGPR is able to capture well the true DMD modes; some mild distortions exist, but the spatial characteristic lengths in each mode are captured well.  As for the interpolated DMD, the modal distortion soon becomes unacceptable once the data percentage is below 40\%.  Furthermore, due to the incorrectly identified frequencies, the mode shape is significantly contaminated by spurious modes such as pure damping modes. This phenomenon appears more extremely for 20\% data case where the characteristic shape of vortex are blurred, and it is almost impossible to compare the mode shapes to the true DMD modes, especially for higher order modes.

\begin{figure}[H]
\centering
\includegraphics[width=.8\textwidth]{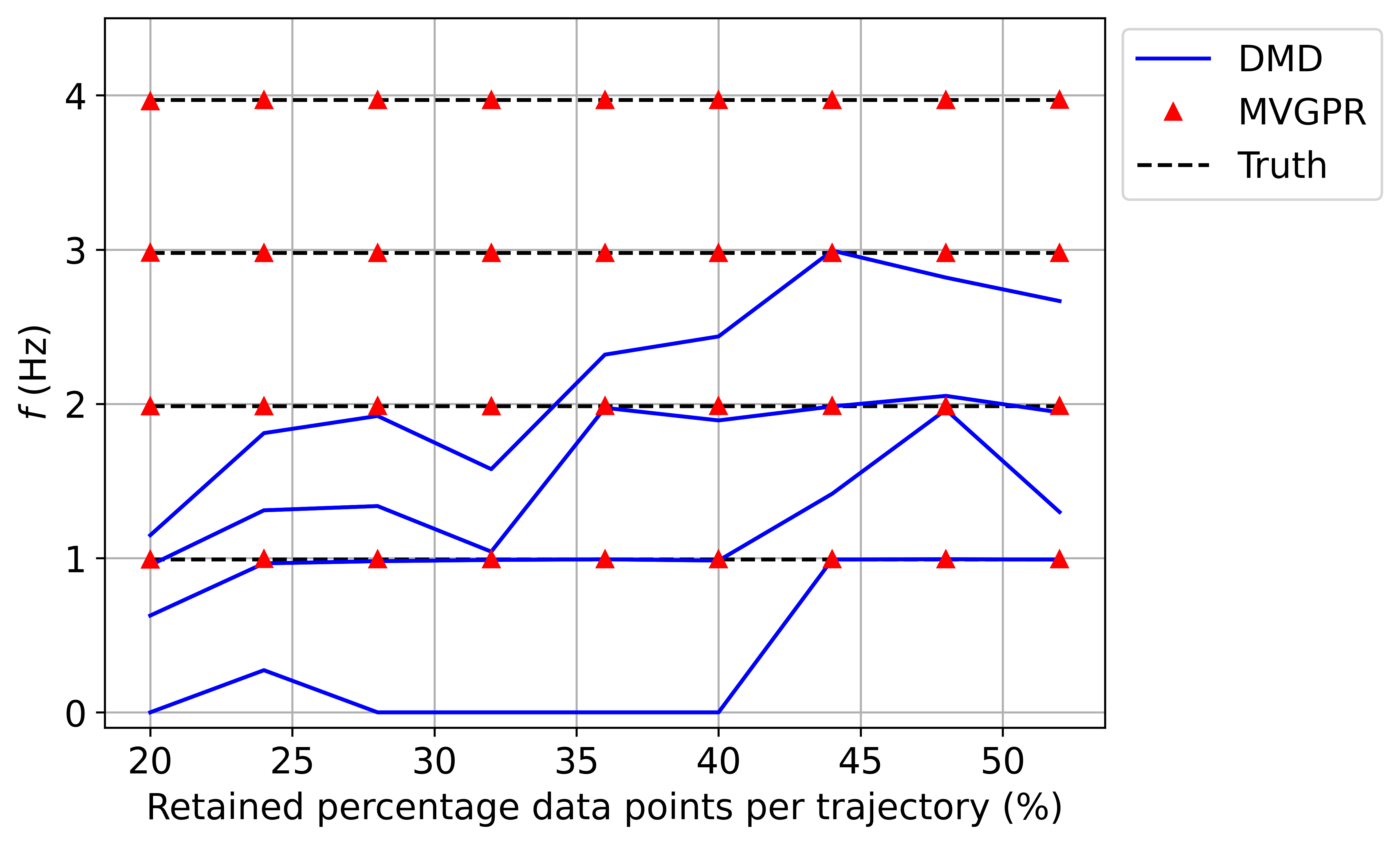}\hspace*{-2cm}
\caption{Identified frequencies of cylinder flow using MVGPR and DMD with cubic interpolation}\label{fig:cylinder_irr_freq_sweep}
\end{figure}

Next, the identified frequencies from both MVGPR and DMD are plotted in Fig. \ref{fig:cylinder_irr_freq_sweep}. It is clear that MVGPR model accurately identifies the frequencies for every percentage considered, while for DMD even 51.34 \% of data points cannot capture the two higher frequencies. As retained data decreases, every identified frequency keeps decreasing, and at 20\%, 3 of them are completely wrong. The error in DMD is attributed to the interpolation procedure that contaminates the frequency components.

\begin{figure}[H]
\centering
\insertfigs{cylinder_true_DMD_modes}{0.248}{True DMD modes}
\insertfigs{cylinder_irr_mvpgr_mode_shape_comp}{0.6}{Identified MVGPR modes} 
\insertfigs{cylinder_true_DMD_modes}{0.248}{True DMD modes}
\insertfigs{cylinder_irr_dmd_mode_shape_comp}{0.6}{Identified DMD modes} 
\caption{Comparison of the identified mode shapes}
\label{fig:dmd_gp_modes_sp}
\end{figure}

\begin{figure}[H]
\centering
\includegraphics[width=.9\textwidth]{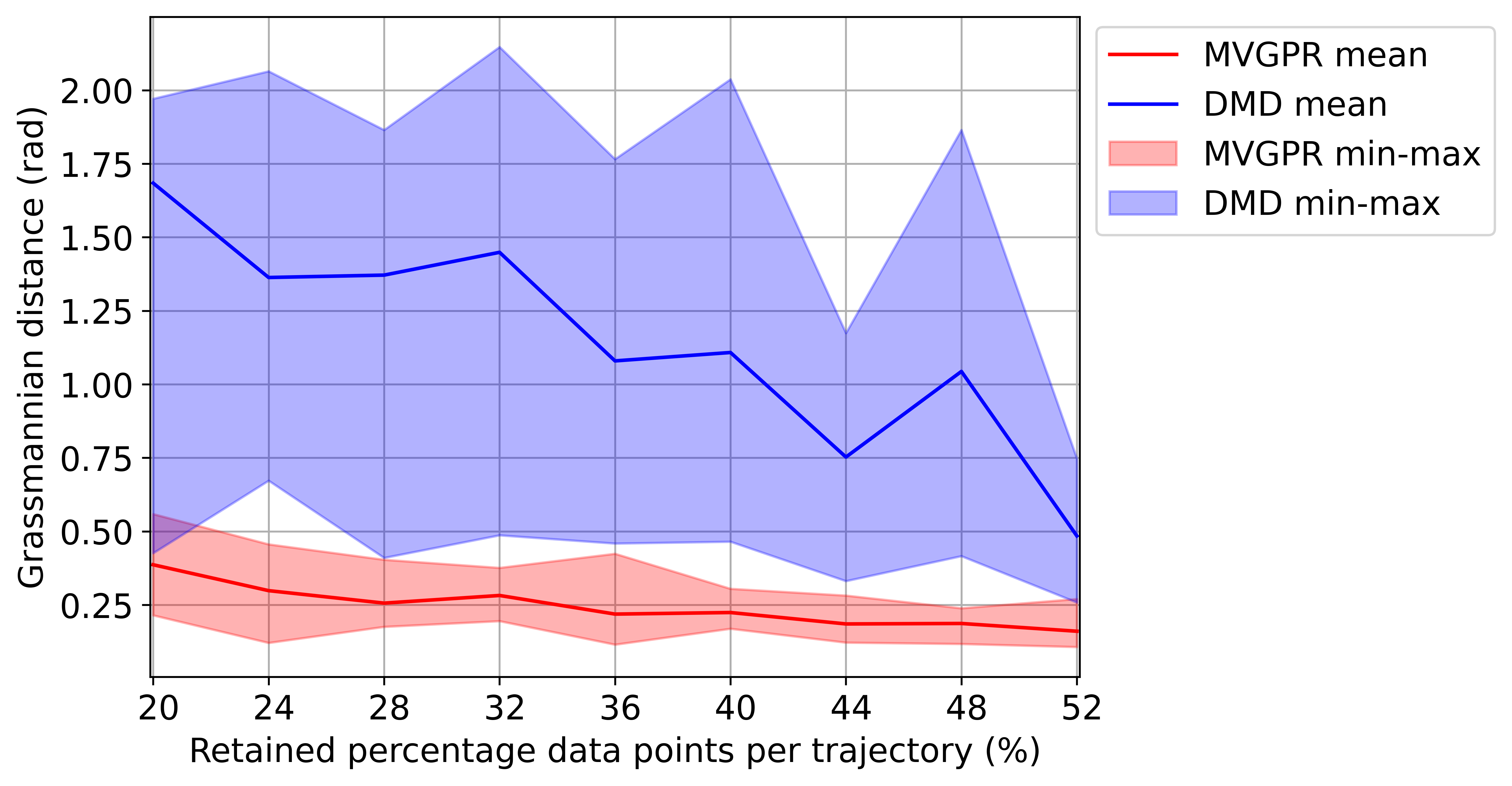}\hspace*{-2.5cm}
\caption{Grassmannian distance for each case considered for MVGPR and DMD}\label{fig:cylinder_irr_sweep}
\end{figure}

Figure \ref{fig:cylinder_irr_measurement} shows the difference between the identified modes and true DMD modes for a range of retained percentages using the concept of Grassmannian distance \cite{ye2016schubert,Sadagopan2022}.   Unlike the inner product metric that compares two sets of vectors in a pairwise manner, the Grassmannian distance directly compares the subspace spanned by two sets of vectors.  Suppose the two sets of $k$ vectors are collected column-wise in two matrices, $\tilde{\vtG}_1$ and $\tilde{\vtG}_2$, the Grassmannian distance is defined as
\begin{equation}
    d(\tilde{\vtG}_1, \tilde{\vtG}_2) = \left( \sum_{i=1}^k (\cos^{-1}\sigma_i)^2 \right)^{1/2}, \quad d \geq 0
\end{equation}
where $\sigma_i$ are the singular values of $\tilde{\vtG}_1^T\tilde{\vtG}_2$.  From the definition, the Grassmannian distance can be viewed as the angle between two subspace. For example, the Grassmannian distance between the two one dimensional vectors is simply the angle between the two vectors. Therefore, when they are pointing the same direction, $d=0$, and becomes larger as one deviates from the other. In Fig. \ref{fig:cylinder_irr_sweep} for each percentage, the training dataset is randomly generated from the original dataset 10 times, and the means and min-max bounds are computed.  Overall, as expected, the distance deviates away from the true DMD for both models as the dataset becomes more sparse.  However, the impact of data sparsity is much more dominant in interpolated DMD than the MVGPR model, both in terms of the mean and the min-max bounds.  Note that on the sparser end of data, the distance of interpolated DMD reaches $\sim$1.7 rad, which indicates that some identified modes are orthogonal to the true modes and hence totally off.  Furthermore, the wide min-max bounds of interpolated DMD mean that the method is highly sensitive to the choice of data points and vulnerable to missing data.  On the contrary, the MVGPR model is able to consistently produce relatively accurate modes throughout the retained percentages.  Again, such superiority of MVGPR is attributed to the inductive bias endowed by the periodic kernel structure, that extracts more dynamical information from the sparse dataset.

\subsection{Problem 2: Synthesized Flow Data}
In this example, we will demonstrate the validity of MVGPR when there exits multiple modes for each frequency in the data. For this purpose, we considered following form of system dynamics where for each particular frequency, two different first order oscillatory responses are combined with amplitudes given by the modes.

\begin{equation}
\vg(t, x, y) = \sum_{i=1}^4 M_{ir} \cos(\omega_i(t + \alpha_i)) + M_{ic}\sin(\omega_i(t + \alpha_i))
\end{equation}
where the mode shapes are arbitrarily chosen.
\begin{equation}
\begin{split} 
M_{1r} &= e^{(x-0.5)^2 + (y-0.5)^2},\\
M_{2r} &= \sin(2x)\sin(2y), \\
M_{3r} &= \sin(4x)\sin(4y)M_{1r}, \\
M_{4r} &= \sin(4x)\sin(2y),
\end{split}
\quad
\begin{split}
M_{1c} &= e^{(x-0.2)^2 + (y-0.2)^2}, \\
M_{2c} &= \cos(2x)\cos(2y), \\
M_{3c} &= \cos(4x)\cos(4y)e^{(x-0.2)^2 + (y-0.2)^2}, \\
M_{4c} &= \cos(4x)\cos(2y)
\end{split}
\end{equation}

There are multiple latent processes that evolve in time with the relative phase angles, $\alpha_{i}$ which are sampled randomly using Gaussian distribution. The two frequencies of the system are $f_1 = 3.1$ Hz and $f_2=5.2$ Hz respectively, and $\omega_i = 2 \pi f_i$. It should be noted that we need at least two different realizations, otherwise the two mixed modes are not mathematically distinguishable. Therefore, it is required to employ the second version of MVGPR to account for the multiple realizations. In general, more trajectories with different relative phases between the two harmonics gives more information to the model and hence can facilitate the learning process. In this problem, 5 different realizations are used. For MVGPR model training, the initial guess for the frequencies of the Cosine kernels is $7 \%$ deviated from the true values.  One can usually obtain the guessed values via Fast Fourier Transform.  In this case the accuracy of the identified frequency is usually better than $7\%$.  The weights of LMC and linear kernels, we randomly assign the values using Gaussian distribution.

\subsubsection{Full Measurement Data}

\begin{figure}[H]
\centering
\includegraphics[width=.8\textwidth]{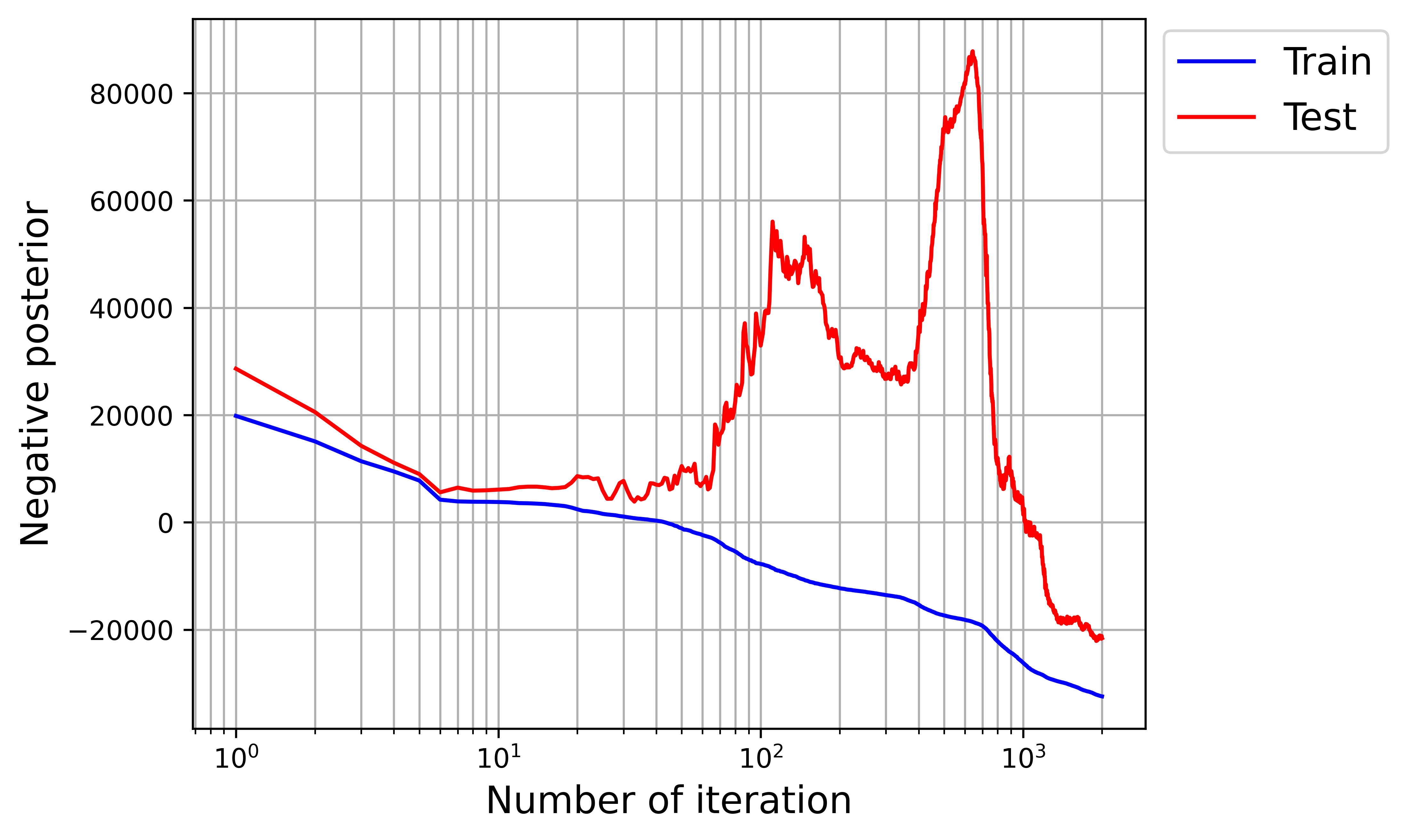}\hspace*{-2cm}
\caption{Negative posterior value with 2k iterations}\label{fig:syn_neg_post}
\end{figure}

We first investigate the problem without temporal irregularity to verify the performance of mode identification compared to SPOD and DMD methods. As there are 8 modes within the data, the POD provides 8 coordinates. For model training, 8 cosine kernels are used, but only 4 of them are trainable and the other 4 are the replicated for conjugate pair. The orthonormal regularization factor, $\lambda_{\cR}$ is set to zero as we are only interested in the subspace of the modes rather than the individual mode shapes in this particular problem. The training takes approximately 21 minutes for 2k iterations.
In Fig. \ref{fig:syn_neg_post}, the negative posterior is computed for each iteration of training. It does not converge with 2k iterations, but for modal analysis purpose, this provides sufficiently accurate result. 

\begin{figure}[H]
\centering
\includegraphics[width=\textwidth]{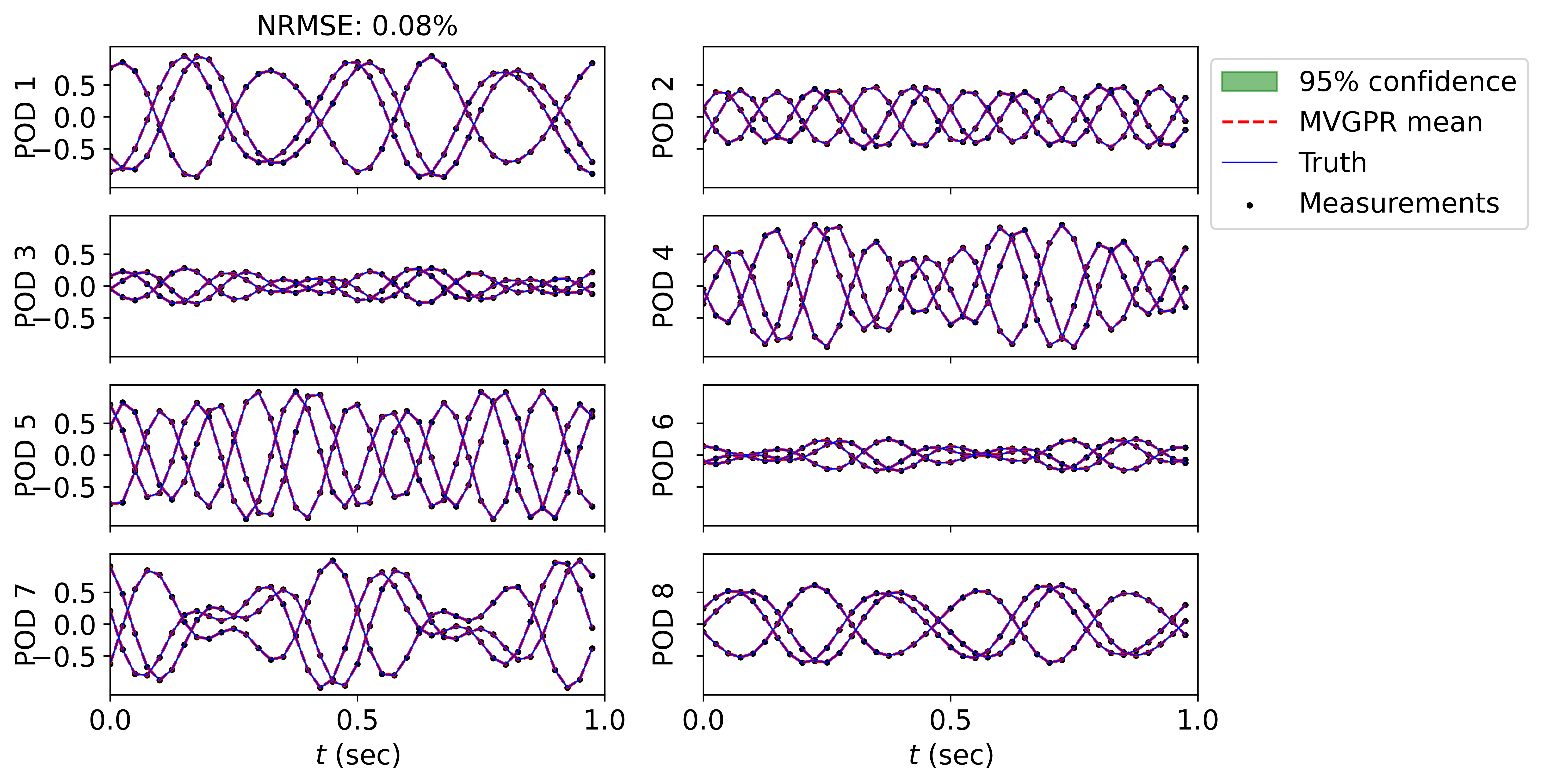}\hspace*{-2cm}
\caption{MVGPR mean and standard deviation of POD coordinates of training data}\label{fig:syn_gp_response_train}
\end{figure}

\begin{figure}[H]
\centering
\includegraphics[width=\textwidth]{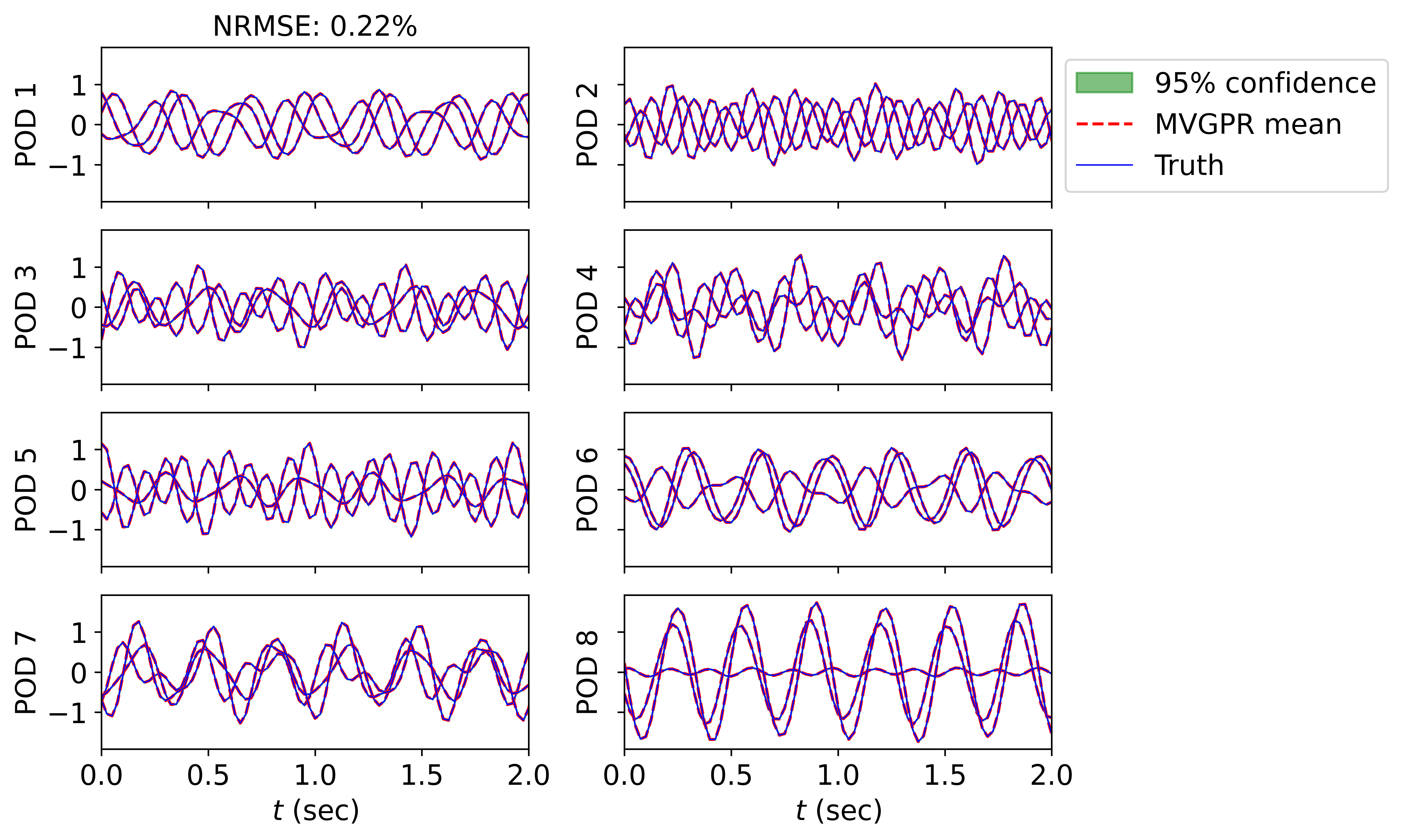}\hspace*{-2cm}
\caption{MVGPR mean and standard deviation of POD coordinates of test data}\label{fig:syn_gp_response_test}
\end{figure}

In Fig. \ref{fig:syn_gp_response_train} and \ref{fig:syn_gp_response_test}, the trajectories used to compute the posterior above are shown. In Fig. \ref{fig:syn_gp_response_train}, three realizations of the system response are plotted   
In Fig. \ref{fig:syn_gp_response_test}, the three test trajectories are shown. Note that in this case, only the initial condition is fed into the MVGPR model, and the trained model can estimate the value at $\tau$ from the initial value given. The NRMSE computed for both training and test data. The calculated error can be further reduced by having more iterations, but this does not change the identified modes and hence is not required for modal analysis.

The modes are identified from each method considered and compared to the true modes used for data generation (Truth). Figure \ref{fig:syn_modes_comp} shows the mode shape comparison. Note that in this case the identified modes are transformed via Grassmannian. One can clearly observe that the all the identified modes are equivalent to the true modes.

\begin{figure}[H]
\centering
\includegraphics[width=.8\textwidth]{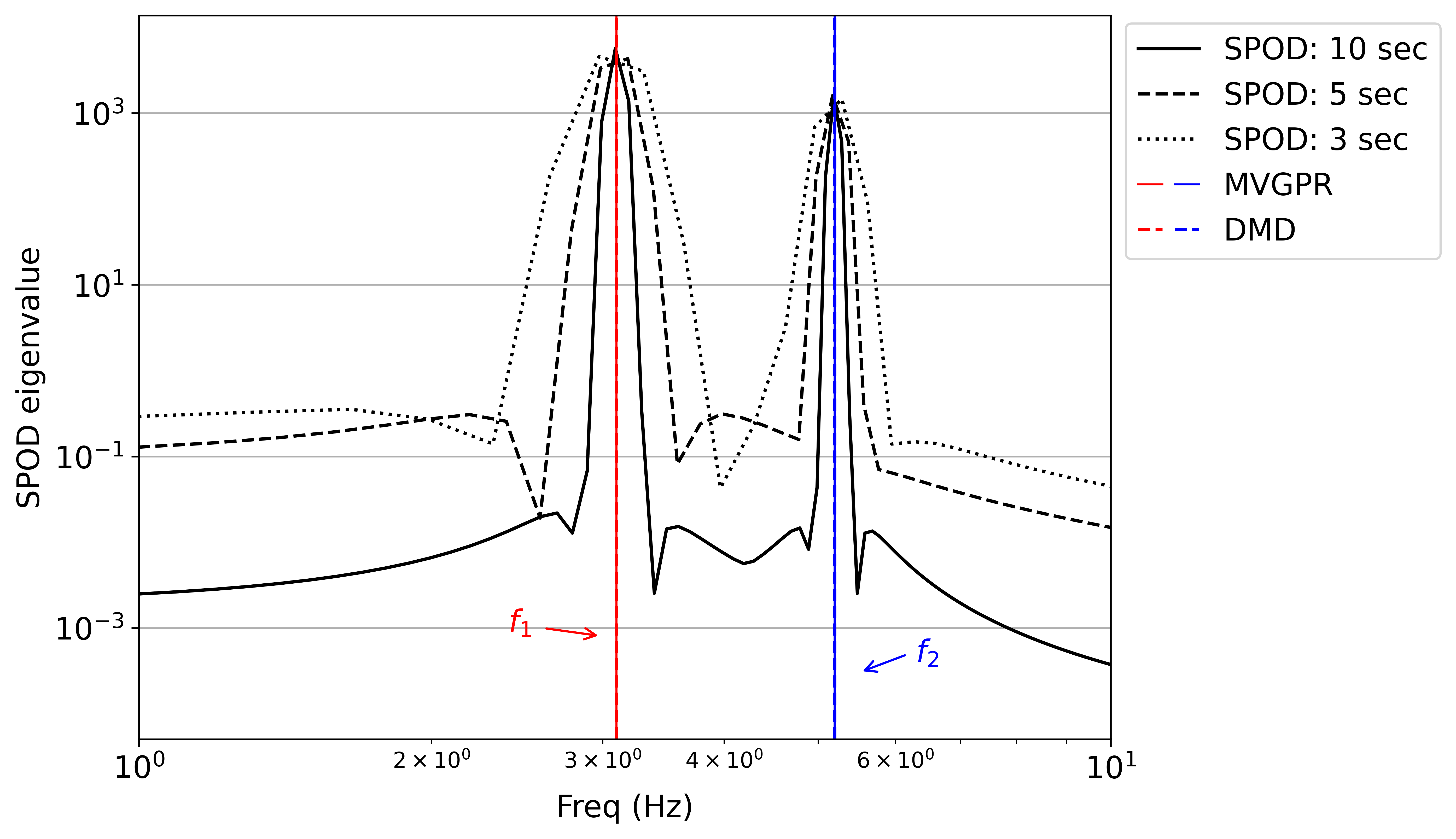}\hspace*{-2cm}
\caption{SPOD eigenvalue plot with three different time span compared to MVGPR and DMD methods}\label{fig:syn_spod_eigs}
\end{figure}

When it comes to the identification of frequencies, it is not difficult to capture the right frequencies for linear systems in general. However, the resolution of the identified frequencies can vary depending on the methods. The most dominant SPOD eigenvalues at different frequencies are computed and are shown in Fig. \ref{fig:syn_spod_eigs}. Here we consider three different time span, 10 sec, 5 sec, and 3 sec. As we decrease the time span, the spectral resolution decreases linearly. For the cases of 5 sec and 3 sec, the SPOD model cannot accurately identify the lowest frequency. However, for MVGPR and DMD, the two frequencies are captured well using the dataset with span of 3 sec. This can be critical if the simulation is expensive to execute, and in general, one cannot be completely certain about the appropriate length of the simulation.

\begin{figure}[H]
\centering
\insertfigs{syn_mode_comparison_freq1}{0.45}{First frequency component, $f_1$}
\insertfigs{syn_mode_comparison_freq2}{0.45}{Second frequency component, $f_2$}
\caption{Mode shape comparison between modes used for data generation (Truth) and the ones identified by MVGPR, DMD, and SPOD respectively}
\label{fig:syn_modes_comp}
\end{figure}

\subsubsection{Temporally Irregular Measurement}

\begin{figure}[H]
\centering
\includegraphics[width=\textwidth]{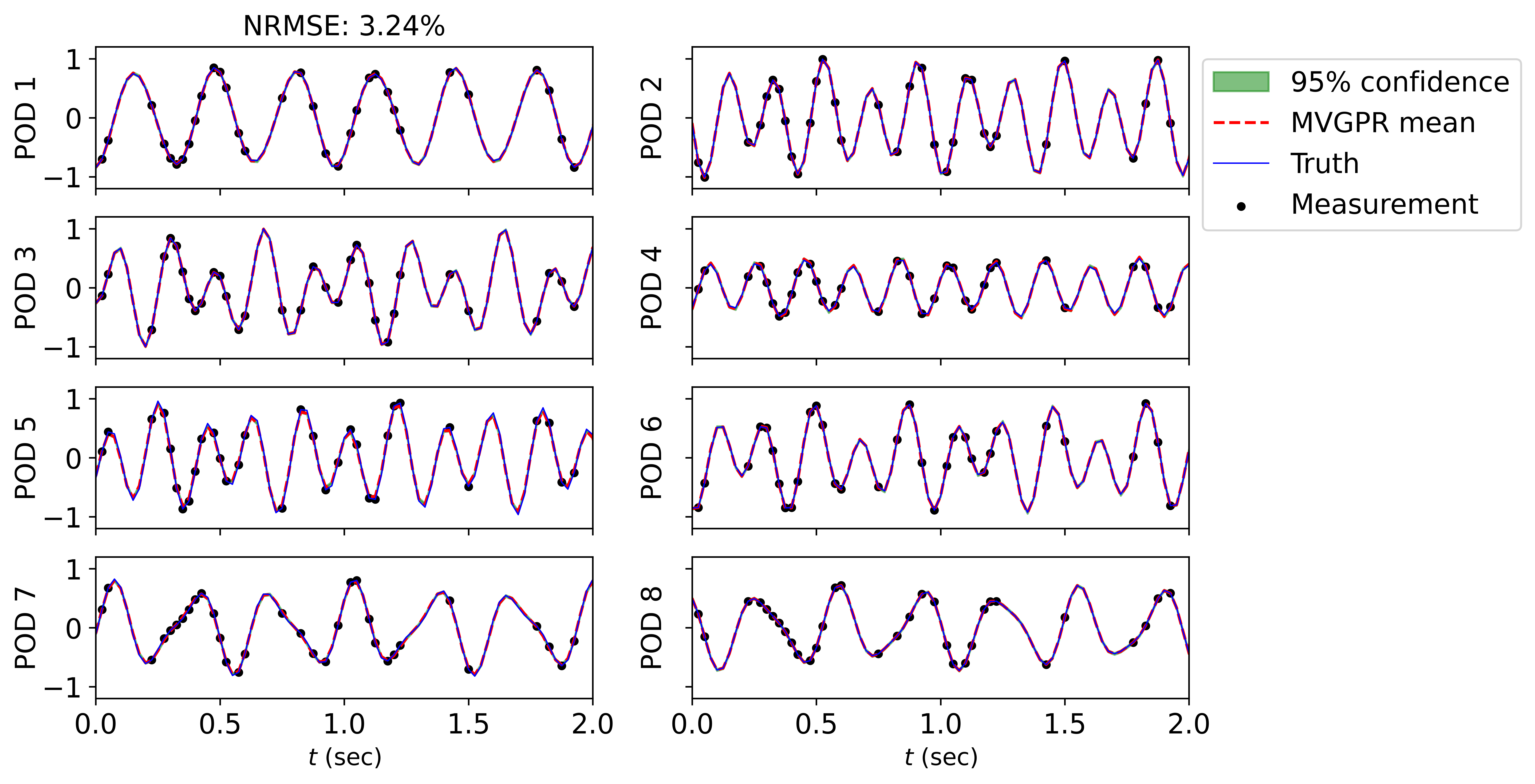}\hspace*{-2cm}
\caption{MVGPR mean and standard deviation of POD coordinates of irregular measurement}\label{fig:syn_gp_irr_response_train}
\end{figure}

\begin{figure}[H]
\centering
\includegraphics[width=.9\textwidth]{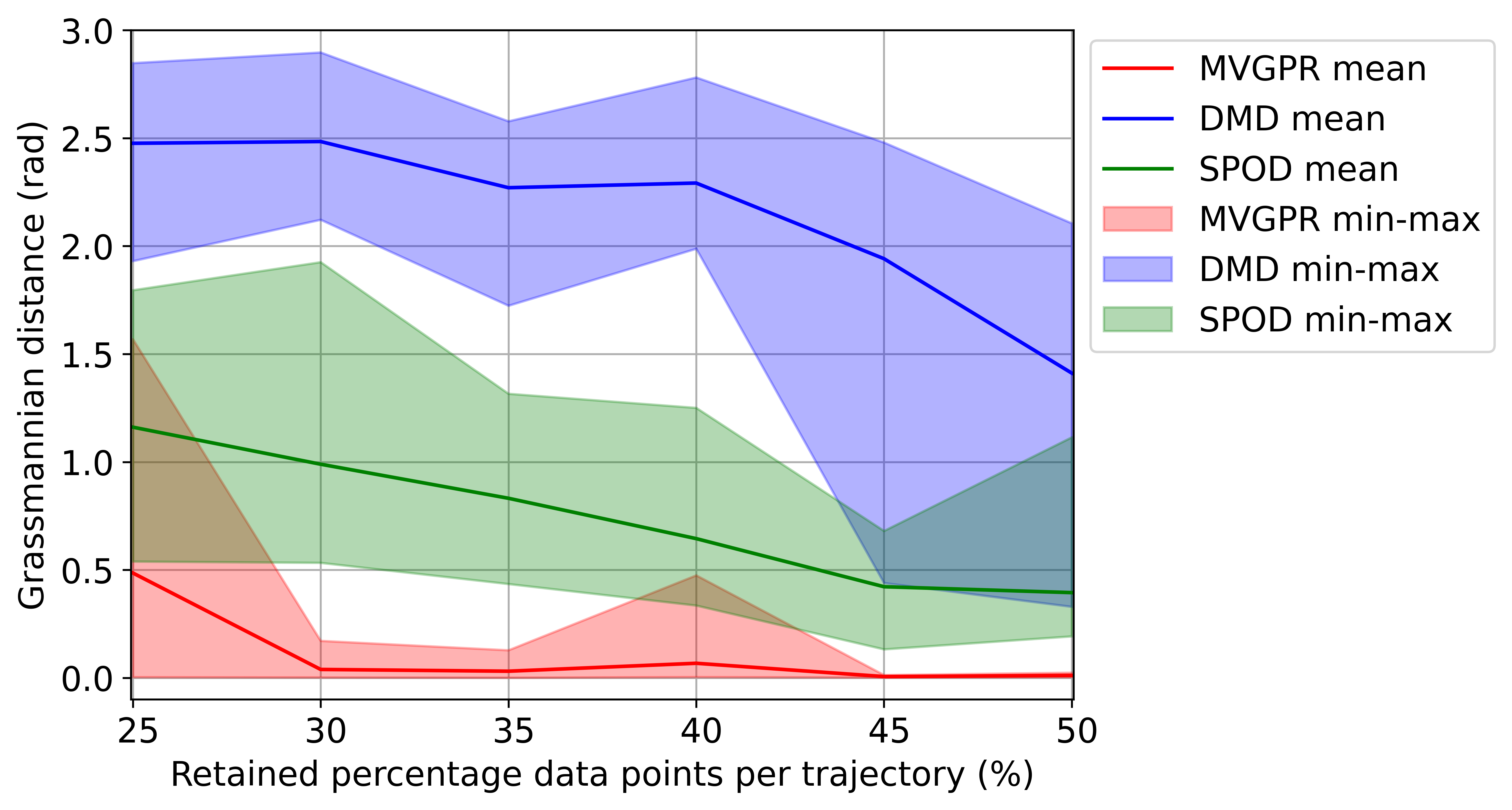}\hspace*{-2cm}
\caption{Grassmannian distance for each case considered for MVGPR, DMD, and SPOD}\label{fig:syn_irr_GM_dist}
\end{figure}

\begin{figure}[H]
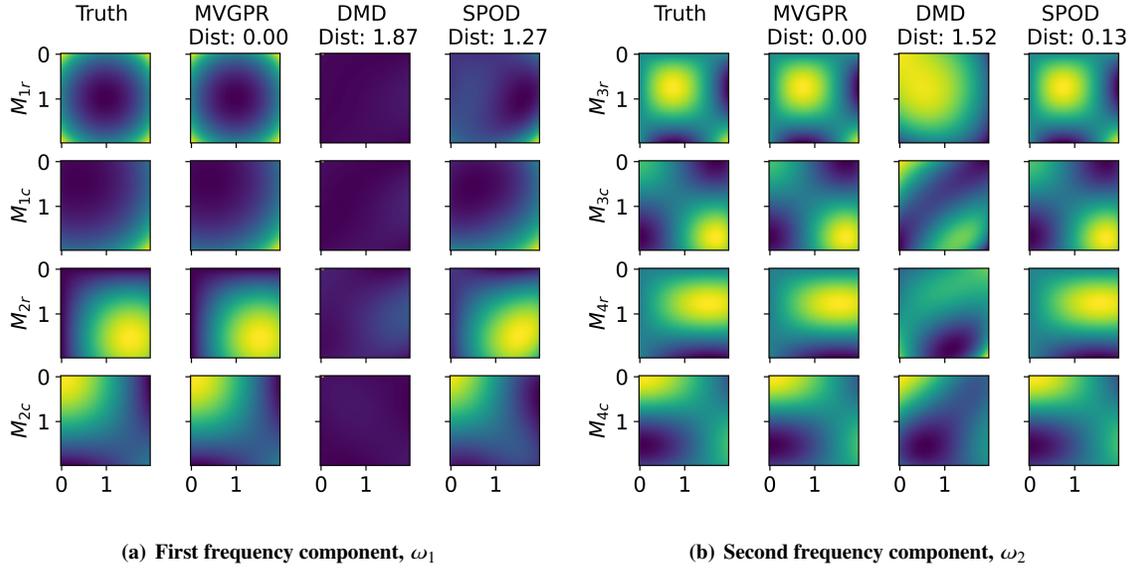

\centering
\insertfigs{syn_irr_mode_comparison_freq1}{0.45}{First frequency component, $\omega_1$}
\insertfigs{syn_irr_mode_comparison_freq2}{0.45}{Second frequency component, $\omega_2$}
\caption{Mode shape comparison between modes used for data generation (Truth) and the ones identified by MVGPR, DMD, and SPOD using 30\% retained percentage data points}
\label{fig:irr_dmd_gp_modes}
\end{figure}

Similar to the cylinder vortex flow case, we consider temporally irregular measurement here. In this section, The retained data points range from 25\% to 50\% with 5\% step size, which results in 6 cases in total. The result of the MVGPR model response using 40\% data points is shown in Fig. \ref{fig:syn_gp_irr_response_train}. Even though there are sparser and irregular measurement, the trained MVGPR can successfully interpolate the data.  When Grassmannian distance is calculated, modes of each frequency are compared and two distances are obtained from $\omega_1$ and $\omega_2$, and the Euclidean distance is calculated from the two. For each case considered, 10 different sample trajectories are realized and used for MVGPR, DMD, and SPOD. In Fig. \ref{fig:syn_irr_GM_dist}, the calculated Grassmannian distance is shown. The mean of 10 cases is connected with solid line and the minimum and maximum bound is represented with transparent area. From 50\% to 25\%, one can clearly observe the overall increasing trend for DMD and SPOD methods. Overall, SPOD outperforms DMD, and MVGPR outperforms SPOD.
In SPOD analysis, the modes corresponding to the true frequencies are used to compute the Grassmannian distance. However, when it comes to DMD, the computed spectrum is contaminated by the interpolation and hence result in artificial modes. For MVGPR, the mean Grassmannian distance is close to zero if retained percentage data points is larger than $25 \%$. The min-max bound of MVGPR is narrower than other methods. There is a small peak at $40 \%$ retained percentage data points. 
Figure \ref{fig:irr_dmd_gp_modes} shows the mode shapes of all three methods for one of the realizations of 30\% retained percentage data points case. The Grassmannian distance calculated from each method is also displayed. The distance of MVGPR model is zero while the other models show conspicuous non-zero distance. DMD model captures significantly inaccurate mode shapes, especially for $\omega_1$. This is because interpolated data points introduce non-physical damping modes, which is more critical to the lowest frequency component. For the SPOD model, even though the total distance is much larger than MVGPR model, most distance is attributed to $\omega_1$, and the distance for $\omega_2$ component is only about 0.13. 

\subsection{Problem 3: Pitching \& Plunging Flat-plate}

This section aims at analyzing more realistic problem which is pitching and plunging flat plate. The pitching and plunging motion of the plate provides rich modal behavior of the system. As emphasized in the previous problem, it is required to have multiple realizations with different phase of motion. In this case, the phase of pitching motion needs to vary with respect to the phase of the plunging motion. Direct numerical simulations of the incompressible Navier–Stokes equations are performed using an immersed boundary projection method \cite{taira2007immersed, colonius2008fast}. In Fig. \ref{fig:ppplate}, the 2D computational domain (a) is shown along with the illustration of pitching (b) and plunging (c) motion. The domain is defined by $600 \times 300$ spatial grid points. For every solution, the Reynolds number is set to 100.

\begin{figure}[H]
\centering
\insertfigs{pics/pitchpluge_setup}{0.9}{Computational domain of flat plate in pitching and plunging motion}
\insertfigs{pics/pitchpluge_setup_pitch}{0.45}{Pitching motion setup}
\insertfigs{pics/pitchpluge_setup_plunge}{0.45}{Plunging motion setup}
\caption{2D computational domain and pitching and plunging motion setup}\label{fig:ppplate}
\end{figure}

The pitching and plunging behaviors can be described by Eq. \eqref{eqn:pp_motion}. In the analysis, the reference term of plunging motion is $h_0 = 0$ so that the mean vertical position of the airfoil stays the center of the domain. When multiple solutions are generated pitching angle is $\phi_{\alpha} = 0$ while only $\phi_{h}$ is changed. The reference pitching angle is set to $\alpha_{0} = 0.5236$ or $30 \deg$. The amplitudes of the pitching and plunging motions are $A_{\alpha}=0.1, \; A_{h} = 0.2$. $\omega_p = 2\pi f_p c/U_{\infty}$ is the non-dimensional forcing frequency, and $c$ is the airfoil chord length, and $U_{\infty}$ is the freestream velocity. This non-dimensional frequency is same for pitching and plunging motion, and it is obtained based on the static airfoil simulation data which is approximately $f_p=0.2549$. The pivot of the pitching motion is located at the quarter of the chord length from the leading edge of the plate. When the pitching and plunging frequencies coincide with the frequency of the natural vortex, the system is expected to be synchronized, so that the natural vortex frequency locks on the frequency of the pitching and plunging motion. Provided this prior knowledge, one can infer that there may exist multiple modes that share the same frequency component. In this case, they will be the natural vortex mode, pitching and plunging modes.

\begin{equation}\begin{split}\label{eqn:pp_motion}
    \alpha(t) &= A_{\alpha} \sin(\omega_p(t + \phi_{\alpha})) + \alpha_0 \\
    h(t) &= A_h \sin(\omega_p(t + \phi_h)) + h_0
\end{split}
\end{equation}

Four realizations are sampled with different plunge phase, $\phi_h = \{0.0, 1.0, 1.8, 2.3\}$. The simulation time span is $32 c/U_{\infty}$, where the step size is $\Delta t = 0.01c/U_{\infty}$. The data is collected at every $10 \Delta t$, which gives 321 temporal steps.  To keep only the post-transient response, first 200 steps are ignored in the following modal analyses. For convenience, we compare the mode shapes corresponding to the lowest frequency. 

\subsubsection{Full Measurement Data}

As mentioned earlier, it is expected that the system has 3 distinct modes for each frequency. In this section, we verify this by DMD, SPOD, and MVGPR methods and compare the difference in performance.  As a first case, DMD method is considered. For this, POD coordinates with 99\% singular values are retained.  The percentage that is less than 99\% results in incomplete number of modes.  The discrete-time DMD eigenvalues are calculated and shown in Fig. \ref{fig:ppdmdeigs}.  The DMD model with a single trajectory and ensemble DMD with multiple trajectories are considered.  Similar to the cylinder vortex flow considered earlier, the discrete spectrum of the system is formed such that the higher frequencies are integer multiples of the lowest frequency. The DMD with single trajectory is certainly not able to capture the multiple modes existing in each frequency component. To capture the 3 modes per frequency, at least 3 realizations are needed in the data, and since 5 realizations are considered in ensemble DMD, it can successfully obtain all 3 modes.  In Fig. \ref{fig:ppdmdeigs} (b), another interesting observation is that the DMD model with single trajectory captures the eigenvalues on the unit circle, but when ensemble DMD captures the multiple modes, their spectra is polluted. This yields non-physical positive damping, which is known as spectral pollution in literature. This pollution gets severe for higher frequency. 

\begin{figure}[H]
\centering
\includegraphics[width=.7\textwidth]{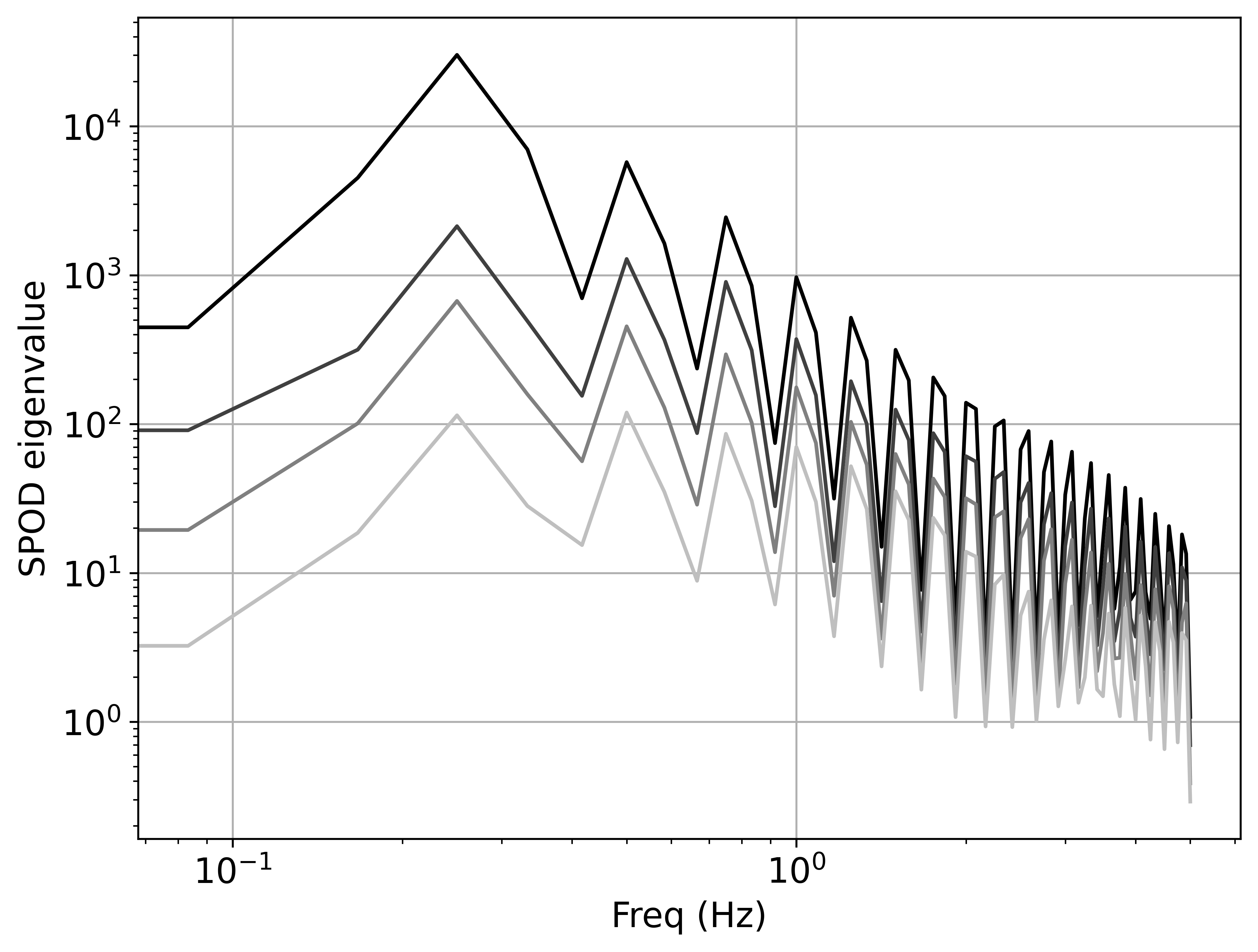}
\caption{SPOD eigenvalues for a range of frequency}\label{fig:ppspodeigs}
\end{figure}

As a next case, SPOD method is considered. In this analysis, each realization forms a single block matrix without overlap between the blocks, which gives 4 blocks of dataset in total. In Fig. \ref{fig:ppspodeigs}, the SPOD eigenvalues for a range of frequencies are shown. As similar to the DMD method, the figure shows dominant frequency components which are the integer multiples of the lowest frequency. One can clearly see that the lowest frequency has the largest eigenvalue, and the peak eigenvalue monotonically decreases as the frequency increases. For every frequency, it calculates 4 different modes from 4 blocks of dataset. The top one with the darkest line corresponds to the largest eigenvalue. The second and third modes have eigenvalues whose order of magnitude is slightly less than the largest one. From 4th mode, this order of magnitude is less than $1\%$ or $2\%$ of the dominant frequency and hence can be ignored. Therefore, the first 3 modes for each frequency and their complex conjugate form the complete bases of the system. As a reference, the cumulative normalized SPDO eigenvalues at the lowest frequency are  0.912, 0.976, 0.996, 1.0.

In the MVGPR model, it is possible to make the kernel parameters shared for different kernel functions, which allows to learn the same frequency while capturing different modes associated with that frequency component. As SPOD first captures 4 modes out of 4 blocks of dataset and truncate the modes using the calculated eigenvalues, the MVGPR model first tries to capture 8 modes including real and imaginary parts and truncate the modes at some threshold based on the calculated $\tilde{\lambda}_j$. In Table. \ref{tab:ppfreqall}, the actual identified frequencies for the first 2 lowest frequencies are shown. For DMD, there are 3 different values where each corresponds to the different modes. However, the SPOD model identifies the three equivalent frequencies. The difference between the two methods is obvious especially for $f_1$, which is mainly attributed to the spectral discretization for FFT calculation in SPOD.

\begin{figure}[H]
\centering
\insertfigs{pics/pitchpluge_full_dmd_deigs}{0.46}{Eigenvalues with unit circle}
\insertfigs{pics/pitchpluge_full_dmd_deigs_zoomin}{0.46}{Zoomed in eigenvalues for the first two lowest frequencies}
\caption{Discrete-time eigenvalues obtained from DMD with single trajectory and ensemble DMD with multiple trajectories}\label{fig:ppdmdeigs}
\end{figure}

\begin{table}[H]
\caption{ Identified frequency from DMD, SPOD, and MVGPR}\label{tab:ppfreqall}
\centering
\begin{tabular}{p{1cm}cccc}
\hline
Frequency & DMD & SPOD & MVGPR  \\
\hline
$f_1$  & (0.2549, 0.2556, 0.2549) & 0.25 & 0.2549\\
$f_2$ & (0.5100, 0.5097, 0.5099) & 0.5  & 0.510\\ 
\hline
\end{tabular}
\end{table}

\begin{figure}[H]
\centering
\includegraphics[width=.6\textwidth]{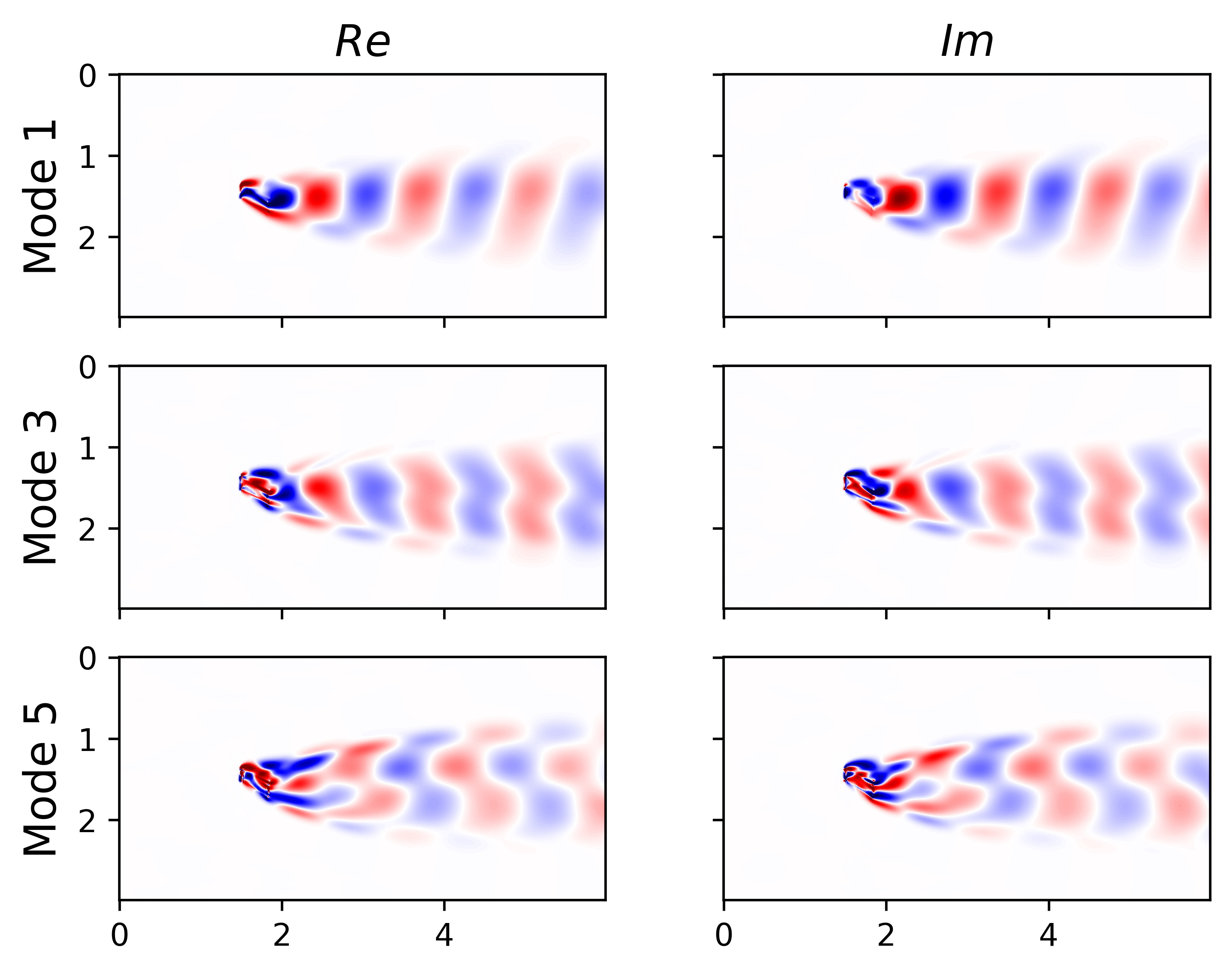}
\caption{Three DMD modes at $f_1$}\label{fig:ppdmdf1}
\end{figure}

In Fig. \ref{fig:ppdmdf1}, three DMD modes at frequency $f_1$ are displayed. Note that the order of the modes has nothing to do with their degree of contribution to the system as DMD has no capability of identifying hierarchical structure of the modes. The odd numbered modes are complex-conjugate of even numbered modes, so only odd numbered modes are displayed. In Fig. \ref{fig:ppdmdf1}, the mode 1 shows the natural vortex mode, and the mode 3 and 5 show the vortex due to pitching and plunging motion of the plate. 

In Fig. \ref{fig:ppSpodGpModes} three SPOD modes and six MVGPR modes at $f_1$ are displayed. The modes are listed from the most dominant to the least dominant based on the calculated SPOD eigenvalues and MVGPR eigenvalues respectively, which are displayed in Table. \ref{tab:ppeigs}. While the calculated SPOD modes are similar to the DMD modes, the mode 2 is slightly different. Furthermore, in the conventional DMD algorithm, there is no mathematical way of identifying the rank of the modes. Therefore, mode-wise comparison is, in general, challenging for complex flows. From the 3 SPOD modes, 3 Grassmannian distances between SPOD and MVGPR modes are calculated, and they are 0.15, 0.61, 0.66 respectively. The identified MVGPR modes seem similar to the SPOD modes, but there is certainly difference between the two, especially for the mode 3 to 6. During training, capturing mode 1 and 2 are not so challenging, but it is difficult to capture more accurate SPOD modes for other modes. In fact, even with the orthonormal regularization, it is still possible for the modes to be mixed. In addition, mode 5 and 6 are not precisely 90 degrees phase off.

\begin{figure}[H]
\centering
\insertfigs{pics/pitchpluge_full_spod_mode_f1}{0.6}{First three dominant SPOD modes}
\insertfigs{pics/pitchpluge_full_mvgpr_mode_f1}{0.6}{First six dominant MVGPR modes}
\caption{Identified modes from SPOD and MVGPR at $f_1$}\label{fig:ppSpodGpModes}
\end{figure}

\begin{table}[H]
\caption{ The normalized eigenvalues calculated from SPOD and MVGPR}\label{tab:ppeigs}
\centering
\begin{tabular}{p{3cm}cccc}
\hline
Model \textbackslash \; Mode (j) & 1 & 1,2 & 1-3 & 1-4 \\
\hline
SPOD ($\lambda_j/\lambda_4$) & 0.912 & 0.976 & 0.996 & 1.0\\
MVGPR ($\tilde{\lambda}_j/\tilde{\lambda}_4$) & 0.815 & 0.916  & 0.975 & 1.0\\ 
\hline
\end{tabular}
\end{table}

\subsubsection{Temporally Irregular Measurement}
Lastly, we consider temporally irregular measurement of pitching and plunging plate system. For this, we randomly select 30 data points out of 121 data points for each realization, which gives 120 data points in total. DMD is not considered in this case because its vulnerability to interpolation is already demonstrated in the previous two cases. Here, we specifically focus on the comparison between the SPOD and MVGPR. For SPOD, the selected data points are used for cubic interpolation to recover the original dataset, which is subsequently used for SPOD analysis. On the other hand, MVGPR model is trained directly on the irregular dataset. In Fig. \ref{fig:ppIrrSpodGpModes}, the identified SPOD and MVGPR modes are shown. For SPOD, when only 30 data points are used, the mode shape so contaminated that in Fig. \ref{fig:ppIrrSpodGpModes} (a), it seems that SPOD produces another dominant mode as a artifact of the interpolation, and the rest of the modes are pushed down by one. In fact, this does not occur in the case with 36 data points per trajectory which will be further discussed next.

\begin{figure}[H]
\centering
\includegraphics[width=.9\textwidth]{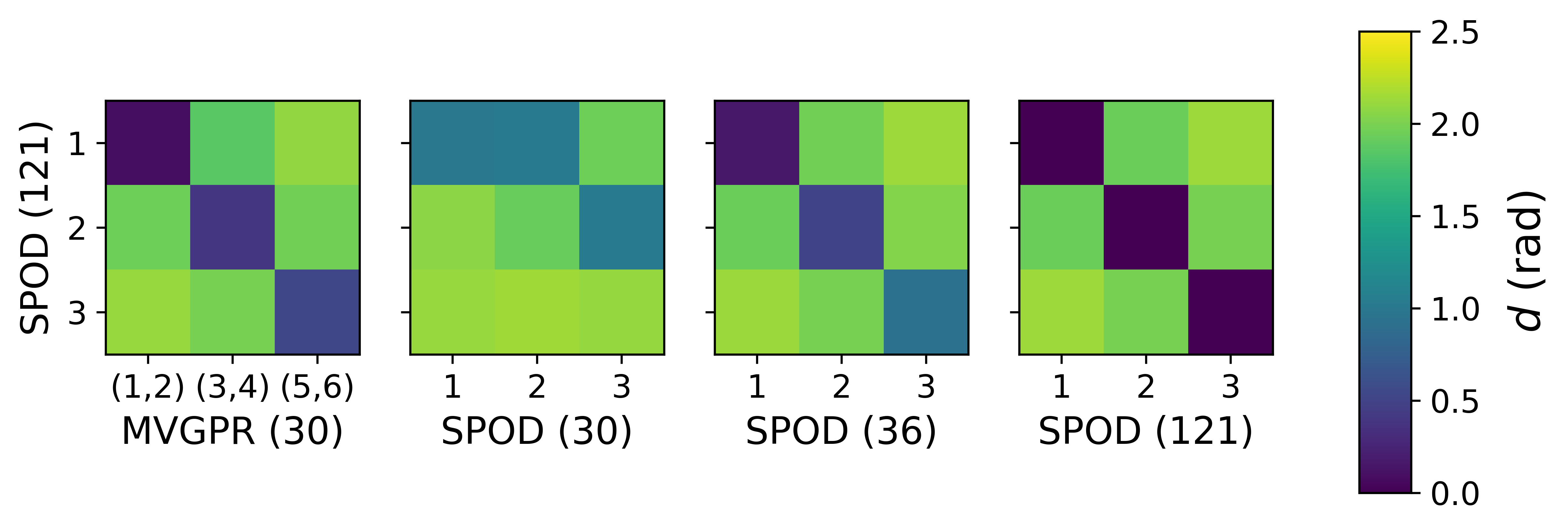}
\caption{Quantitative mode shape comparison}\label{fig:ppIrrCmp}
\end{figure}

The calculated pair-wise mode shape comparison is shown in Fig. \ref{fig:ppIrrCmp}. The number in parenthesis denotes the number of data points per trajectory. Each case is compared to the SPOD modes identified using the dataset with 121 data points per trajectory which is also shown on the right side of the plot as a reference. Another case with 36 data points per trajectory is also shown. As mentioned earlier, from SPOD (36) to SPOD (30), the performance becomes significantly worse, which is mainly attributed to the artificial modes and misplacement of the other modes. However, MVGPR (30) still outperforms SPOD (36), which demonstrates its superiority. 

\begin{figure}[H]
\centering
\insertfigs{pics/pitchpluge_irr_spod_mode_f1_30}{0.6}{First three dominant SPOD modes}
\insertfigs{pics/pitchpluge_irr_mvgpr_mode_f1}{0.6}{First six dominant MVGPR modes}
\caption{Identified modes from SPOD and MVGPR at $f_1$}\label{fig:ppIrrSpodGpModes}
\end{figure}

\section{Conclusion}\label{sec5}

In this paper, we introduced MVGPR as a novel approach to modal analysis for spatiotemporal data, offering an alternative to traditional methods such as DMD and SPOD.  Our framework established a connection between DMD and MVGPR through linear system identification within the context of Koopman operator theory.  This system identification was further extended to the spectral analysis, which allowed the theoretical linkage between SPOD and MVGPR.  The proposed MVGPR model incorporates linear kernel functions to capture trajectory phase, enabling the prediction of future states from an unseen initial state. In addition, similar to the eigenvalues of SPOD, the variances of the diagonal kernel functions in LMC represent the rank of the identified modes. In fact, when orthonormal regularization of the modes is imposed during training, the MVGPR model identifies the SPOD modes with a reasonable rank. 

We verified the superiority of MVGPR over SPOD and DMD in some use cases. First, in the latter two methods, the identified spectrum could be inaccurate due to either spectral pollution or numerical discretization. However, the MVGPR model can pinpoint the point spectrum using judiciously designed kernel function. Second, it was demonstrated that when data is temporally irregular, MVGPR identifies more accurate modes compared to the DMD and SPOD with interpolated data. This is because the interpolation contaminates the spectral information of the data, which could also contaminate the mode shapes. This effect appeared more conspicuously in DMD modes. 

In this paper, the MVGPR model is developed and demonstrated for stationary flow systems.  However, given the versatility of its kernel design, MVGPR could be extended to tackle more complex dynamics that classical DMD and SPOD methods cannot.  Two examples include application to non-isometric transient dynamics, i.e., those involving damping, and chaotic dynamics with continuous spectrum.  Another direction is to extend the method for more complex data structures, examples include unstructured data defined on, e.g., graphs, and spatially irregular data sets.  Ultimately these capabilities may allow MVGPR to become a more generalized tool for modal analysis of spatiotemporal dynamics.

\bibliography{references}

\begin{thebibliography}{33}
\newcommand{\enquote}[1]{``#1''}
\providecommand{\natexlab}[1]{#1}
\providecommand{\url}[1]{\texttt{#1}}
\providecommand{\urlprefix}{URL }
\expandafter\ifx\csname urlstyle\endcsname\relax
  \providecommand{\doi}[1]{\discretionary{}{}{}https://doi.org/#1}\else
  \providecommand{\doi}[1]{\discretionary{}{}{}\urlstyle{rm}\url{https://doi.org/#1}}\fi

\bibitem[{Berkooz et~al.(1993)Berkooz, Holmes, and Lumley}]{berkooz1993proper}
Berkooz, G., Holmes, P., and Lumley, J.~L., \enquote{The proper orthogonal
  decomposition in the analysis of turbulent flows,} \emph{Annual review of
  fluid mechanics}, Vol.~25, No.~1, 1993, pp. 539--575.

\bibitem[{Schmidt and Colonius(2020)}]{schmidt2020guide}
Schmidt, O.~T., and Colonius, T., \enquote{Guide to spectral proper orthogonal
  decomposition,} \emph{AIAA journal}, Vol.~58, No.~3, 2020, pp. 1023--1033.

\bibitem[{Towne et~al.(2018)Towne, Schmidt, and Colonius}]{towne2018spectral}
Towne, A., Schmidt, O.~T., and Colonius, T., \enquote{Spectral proper
  orthogonal decomposition and its relationship to dynamic mode decomposition
  and resolvent analysis,} \emph{Journal of Fluid Mechanics}, Vol. 847, 2018,
  pp. 821--867.

\bibitem[{Schmid et~al.(2011)Schmid, Li, Juniper, and
  Pust}]{schmid2011applications}
Schmid, P.~J., Li, L., Juniper, M.~P., and Pust, O., \enquote{Applications of
  the dynamic mode decomposition,} \emph{Theoretical and Computational Fluid
  Dynamics}, Vol.~25, No.~1, 2011, pp. 249--259.

\bibitem[{Kutz et~al.(2016)Kutz, Brunton, Brunton, and
  Proctor}]{kutz2016dynamic}
Kutz, J.~N., Brunton, S.~L., Brunton, B.~W., and Proctor, J.~L., \emph{Dynamic
  mode decomposition: data-driven modeling of complex systems}, SIAM, 2016.

\bibitem[{Brunton et~al.(2021)Brunton, Budi{\v{s}}i{\'c}, Kaiser, and
  Kutz}]{brunton2021modern}
Brunton, S.~L., Budi{\v{s}}i{\'c}, M., Kaiser, E., and Kutz, J.~N.,
  \enquote{Modern Koopman theory for dynamical systems,} \emph{arXiv preprint
  arXiv:2102.12086}, 2021.

\bibitem[{Taira et~al.(2017)Taira, Brunton, Dawson, Rowley, Colonius, McKeon,
  Schmidt, Gordeyev, Theofilis, and Ukeiley}]{taira2017modal}
Taira, K., Brunton, S.~L., Dawson, S.~T., Rowley, C.~W., Colonius, T., McKeon,
  B.~J., Schmidt, O.~T., Gordeyev, S., Theofilis, V., and Ukeiley, L.~S.,
  \enquote{Modal analysis of fluid flows: An overview,} \emph{AIAA Journal},
  Vol.~55, No.~12, 2017, pp. 4013--4041.

\bibitem[{Taira et~al.(2020)Taira, Hemati, Brunton, Sun, Duraisamy, Bagheri,
  Dawson, and Yeh}]{taira2020modal}
Taira, K., Hemati, M.~S., Brunton, S.~L., Sun, Y., Duraisamy, K., Bagheri, S.,
  Dawson, S.~T., and Yeh, C.-A., \enquote{Modal analysis of fluid flows:
  Applications and outlook,} \emph{AIAA journal}, Vol.~58, No.~3, 2020, pp.
  998--1022.

\bibitem[{Tu(2013)}]{tu2013dynamic}
Tu, J.~H., \enquote{Dynamic mode decomposition: Theory and applications,} Ph.D.
  thesis, Princeton University, 2013.

\bibitem[{Arbabi and Mezi{\'c}(2017)}]{arbabi2017study}
Arbabi, H., and Mezi{\'c}, I., \enquote{Study of dynamics in post-transient
  flows using Koopman mode decomposition,} \emph{Physical Review Fluids},
  Vol.~2, No.~12, 2017, p. 124402.

\bibitem[{Hemati et~al.(2017)Hemati, Rowley, Deem, and
  Cattafesta}]{hemati2017biasing}
Hemati, M.~S., Rowley, C.~W., Deem, E.~A., and Cattafesta, L.~N.,
  \enquote{De-biasing the dynamic mode decomposition for applied Koopman
  spectral analysis of noisy datasets,} \emph{Theoretical and Computational
  Fluid Dynamics}, Vol.~31, 2017, pp. 349--368.

\bibitem[{Colbrook and Townsend(2021)}]{colbrook2021rigorous}
Colbrook, M.~J., and Townsend, A., \enquote{Rigorous data-driven computation of
  spectral properties of Koopman operators for dynamical systems,} \emph{arXiv
  preprint arXiv:2111.14889}, 2021.

\bibitem[{Pan et~al.(2021)Pan, Arnold-Medabalimi, and
  Duraisamy}]{pan2021sparsity}
Pan, S., Arnold-Medabalimi, N., and Duraisamy, K., \enquote{Sparsity-promoting
  algorithms for the discovery of informative Koopman-invariant subspaces,}
  \emph{Journal of Fluid Mechanics}, Vol. 917, 2021, p. A18.

\bibitem[{Takeishi et~al.(2017)Takeishi, Kawahara, Tabei, and
  Yairi}]{takeishi2017bayesian}
Takeishi, N., Kawahara, Y., Tabei, Y., and Yairi, T., \enquote{Bayesian dynamic
  mode decomposition.} \emph{IJCAI}, 2017, pp. 2814--2821.

\bibitem[{Kawashima and Hino(2023)}]{kawashima2023gaussian}
Kawashima, T., and Hino, H., \enquote{Gaussian Process Koopman Mode
  Decomposition,} \emph{Neural Computation}, Vol.~35, No.~1, 2023, pp. 82--103.

\bibitem[{Cardinale et~al.(2022)Cardinale, Brunton, and
  Colonius}]{cardinale2022spectral}
Cardinale, C., Brunton, S., and Colonius, T., \enquote{Spectral Proper
  Orthogonal Decomposition via Dynamic Mode Decomposition for Non-Sequential
  Pairwise Data,} \emph{Bulletin of the American Physical Society}, 2022.

\bibitem[{Henry~de Frahan et~al.(2023)Henry~de Frahan, Rood, Day, Sitaraman,
  Yellapantula, Perry, Grout, Almgren, Zhang, Bell et~al.}]{henry2023pelec}
Henry~de Frahan, M.~T., Rood, J.~S., Day, M.~S., Sitaraman, H., Yellapantula,
  S., Perry, B.~A., Grout, R.~W., Almgren, A., Zhang, W., Bell, J.~B., et~al.,
  \enquote{PeleC: An adaptive mesh refinement solver for compressible reacting
  flows,} \emph{The International Journal of High Performance Computing
  Applications}, Vol.~37, No.~2, 2023, pp. 115--131.

\bibitem[{Edition et~al.(2002)Edition, Papoulis, and
  Pillai}]{edition2002probability}
Edition, F., Papoulis, A., and Pillai, S.~U., \emph{Probability, random
  variables, and stochastic processes}, McGraw-Hill Europe: New York, NY, USA,
  2002.

\bibitem[{Sarkka et~al.(2013)Sarkka, Solin, and
  Hartikainen}]{sarkka2013spatiotemporal}
Sarkka, S., Solin, A., and Hartikainen, J., \enquote{Spatiotemporal learning
  via infinite-dimensional Bayesian filtering and smoothing: A look at Gaussian
  process regression through Kalman filtering,} \emph{IEEE Signal Processing
  Magazine}, Vol.~30, No.~4, 2013, pp. 51--61.

\bibitem[{Sadagopan et~al.(2021)Sadagopan, Huang, Duzel, Martin, and
  Hanquist}]{Sadagopan2021a}
Sadagopan, A., Huang, D., Duzel, U., Martin, L.~E., and Hanquist, K.~M.,
  \enquote{Assessment of high-temperature effects on hypersonic
  aerothermoelastic analysis using multi-fidelity multi-variate surrogates,}
  \emph{AIAA Scitech 2021 Forum}, 2021, p. 1610.

\bibitem[{Huang et~al.(2022)Huang, Sadagopan, D{\"u}zel, and
  Hanquist}]{Sadagopan2022}
Huang, D., Sadagopan, A., D{\"u}zel, {\"U}., and Hanquist, K.~M.,
  \enquote{Study of fluid--thermal--structural interaction in high-temperature
  high-speed flow using multi-fidelity multi-variate surrogates,} \emph{Journal
  of Fluids and Structures}, Vol. 113, 2022, p. 103682.

\bibitem[{Song and Huang(2023)}]{song2023modal}
Song, J., and Huang, D., \enquote{Modal Analysis of Spatiotemporal Data via
  Multi-fidelity Multi-variate Gaussian Processes,} \emph{AIAA AVIATION 2023
  Forum}, 2023, p. 4350.

\bibitem[{Mezi{\'c}(2013)}]{mezic2013analysis}
Mezi{\'c}, I., \enquote{Analysis of fluid flows via spectral properties of the
  Koopman operator,} \emph{Annual Review of Fluid Mechanics}, Vol.~45, 2013,
  pp. 357--378.

\bibitem[{Duke et~al.(2012)Duke, Soria, and Honnery}]{duke2012error}
Duke, D., Soria, J., and Honnery, D., \enquote{An error analysis of the dynamic
  mode decomposition,} \emph{Experiments in fluids}, Vol.~52, 2012, pp.
  529--542.

\bibitem[{Askham and Kutz(2018)}]{askham2018variable}
Askham, T., and Kutz, J.~N., \enquote{Variable projection methods for an
  optimized dynamic mode decomposition,} \emph{SIAM Journal on Applied
  Dynamical Systems}, Vol.~17, No.~1, 2018, pp. 380--416.

\bibitem[{Rasmussen and Williams(2006)}]{Rasmussen2006}
Rasmussen, C.~E., and Williams, C. K.~I., \emph{Gaussian {Processes} for
  {Machine} {Learning}}, The MIT Press, Cambridge, MA, 2006.
\newblock \doi{10.7551/mitpress/3206.001.0001}.

\bibitem[{Forrester et~al.(2008)Forrester, Sóbester, and
  Keane}]{Forrester2008}
Forrester, A., Sóbester, A., and Keane, A., \emph{Engineering Design Via
  Surrogate Modelling: A Practical Guide}, John Wiley \& Sons, Hoboken, NJ,
  2008.
\newblock \doi{10.1002/9780470770801}.

\bibitem[{Parussini et~al.(2017)Parussini, Venturi, Perdikaris, and
  Karniadakis}]{Parussini2017}
Parussini, L., Venturi, D., Perdikaris, P., and Karniadakis, G.~E.,
  \enquote{Multi-Fidelity Gaussian Process Regression for Prediction of Random
  Fields,} \emph{Journal of Computational Physics}, Vol. 336, 2017, pp. 36--50.
\newblock \doi{10.1016/j.jcp.2017.01.047}.

\bibitem[{Alvarez et~al.(2011)Alvarez, Rosasco, and Lawrence}]{Alvarez2012}
Alvarez, M.~A., Rosasco, L., and Lawrence, N.~D., \enquote{Kernels for
  vector-valued functions: A review,} , 2011.
\newblock \doi{10.48550/ARXIV.1106.6251}.

\bibitem[{Solin and S{\"a}rkk{\"a}(2014)}]{solin2014explicit}
Solin, A., and S{\"a}rkk{\"a}, S., \enquote{Explicit link between periodic
  covariance functions and state space models,} \emph{Artificial Intelligence
  and Statistics}, PMLR, 2014, pp. 904--912.

\bibitem[{Ye and Lim(2016)}]{ye2016schubert}
Ye, K., and Lim, L.-H., \enquote{Schubert varieties and distances between
  subspaces of different dimensions,} \emph{SIAM Journal on Matrix Analysis and
  Applications}, Vol.~37, No.~3, 2016, pp. 1176--1197.

\bibitem[{Taira and Colonius(2007)}]{taira2007immersed}
Taira, K., and Colonius, T., \enquote{The immersed boundary method: a
  projection approach,} \emph{Journal of Computational Physics}, Vol. 225,
  No.~2, 2007, pp. 2118--2137.

\bibitem[{Colonius and Taira(2008)}]{colonius2008fast}
Colonius, T., and Taira, K., \enquote{A fast immersed boundary method using a
  nullspace approach and multi-domain far-field boundary conditions,}
  \emph{Computer Methods in Applied Mechanics and Engineering}, Vol. 197, No.
  25-28, 2008, pp. 2131--2146.

\end{thebibliography}

\end{document}